\documentclass[parskip=half]{scrartcl}

\title{Scalable Bernoulli Factory MCMC for Intractable Marginalised Posteriors}
\author{
  Timothée Stumpf-Fétizon \\ \href{mailto:timothee.stumpffetizon@unibocconi.it}{\texttt{[timothee.stumpffetizon@unibocconi.it]}} \and
  Flávio B. Gonçalves \\
  \href{mailto:fbgoncalves@ufmg.br}{\texttt{[fbgoncalves@ufmg.br]}}
}

\usepackage[noend]{algpseudocode}
\usepackage{algorithm}
\usepackage{amssymb}
\usepackage{amsthm}
\usepackage[backend=biber, minbibnames=4, maxbibnames=4, uniquelist=false]{biblatex}
\usepackage{booktabs}
\usepackage[colorlinks, citecolor=blue, urlcolor=blue]{hyperref}
\usepackage{listings}
\usepackage{mathtools}
\usepackage{tikz}
\usepackage{todonotes}
\usepackage{xparse}
\usetikzlibrary{shapes.geometric}
\usetikzlibrary{trees}

\NewDocumentCommand\ub{m}{{#1}^{\uparrow}}
\NewDocumentCommand\lb{m}{{#1}^{\downarrow}}

\NewDocumentCommand\class{m}{\mathcal{#1}}
\NewDocumentCommand\meas{m}{\mathbb{#1}}

\DeclarePairedDelimiter{\braces}{\lbrace}{\rbrace}
\DeclarePairedDelimiter{\parens}{\lparen}{\rparen}

\DeclarePairedDelimiter{\abs}{\lvert}{\rvert}

\NewDocumentCommand\dd{g}{\mathop{}\!\mathrm{d} #1}
\NewDocumentCommand\der{gm}{\frac{\dd{} #1}{\dd{} #2}}
\NewDocumentCommand\pder{gm}{\frac{\partial #1}{\partial #2}}
\NewDocumentCommand\op{smO{}O{}g}{
  \IfBooleanTF{#1}{
    \IfNoValueTF{#5}
      {\operatorname*{\mathrm{#2}}_{#3}^{#4}}
      {\operatorname*{\mathrm{#2}}_{#3}^{#4} \left[ #5 \right]}
    }{
    \IfNoValueTF{#5}
      {\operatorname{\mathrm{#2}}_{#3}^{#4}}
      {\operatorname{\mathrm{#2}}_{#3}^{#4} \left[ #5 \right]}
    }
}

\theoremstyle{plain}
\newtheorem{lemma}{Lemma}
\newtheorem{theorem}{Theorem}
\newtheorem{proposition}{Proposition}
\theoremstyle{definition}
\newtheorem{definition}{Definition}
\newtheorem{assumption}{Assumption}
\newtheorem{example}{Example}
\theoremstyle{remark}
\newtheorem{remark}{Remark}

\definecolor{lstbase}{HTML}{555753}
\definecolor{lstkey}{HTML}{5C3566}
\definecolor{lstbg}{HTML}{EEEEEC}
\definecolor{lstcom}{HTML}{888A85}
\definecolor{lstnum}{HTML}{BABDB6}
\definecolor{lstid}{HTML}{555753}
\definecolor{lststr}{HTML}{8F5902}

\let\citep = \cite

\begin{document}
\maketitle

\begin{abstract}
  \emph{Bernoulli factory MCMC} algorithms implement accept-reject Markov chains without explicit computation of acceptance probabilities, and are used to target posterior distributions associated with intractable likelihood models. Intractable likelihoods naturally arise in continuous-time models and mixture distributions, or from the marginalisation of a tractable augmented model. Bernoulli factory MCMC algorithms often mix better than alternatives that target a tractable augmented posterior. However, we show that their computational performance typically deteriorates exponentially with data size, as happens in naively implemented pseudo-marginal algorithms. To address this, we propose a simple divide-and-conquer Bernoulli factory MCMC algorithm and prove that it has polynomial complexity for factorizing models, with the exact degree depending on the existence of efficient unbiased estimators of the intractable likelihood ratio. We demonstrate the effectiveness of our approach with applications to Bayesian inference in two intractable likelihood models, and observe respective polynomial cost of degree 1.2 and 1 in the data size.
  \par\vskip\baselineskip\noindent
  \textbf{Code:} \url{https://github.com/timsf/scale-bernoulli-mcmc}
  \par\vskip\baselineskip\noindent
  \par\vskip\baselineskip\noindent
  \textbf{MSC2020 subject classifications:} Primary 62-08; Secondary 62F15
\end{abstract}

\section{Introduction to Bernoulli Factory MCMC}
\label{sec:bmcmc}

Markov chain Monte Carlo (MCMC) algorithms are widely used to sample from high-dimensional distributions whose densities are computable up to a normalizing constant. This problem often arises in Bayesian statistics when inferring parameters $\theta$ in some parameter space $\class{T} \subseteq \mathbb{R}^{d}$ from data $y$. Models frequently include a nuisance variable $z$ which affords various conveniences, e.g.\ factorization of the complete likelihood $\pi(y, z | \theta)$, or the existence of bounded estimators. Where the likelihood $\pi(y | \theta)$ already provides those, we can proceed without nuisance variables. We further presume that $\pi(z | \theta)$ is a density with respect to some dominating measure $\meas{M}(\dd{z})$, and set some prior density $\pi(\theta)$ with respect to Lebesgue measure. On that basis, the Bayesian posterior distribution is given by
\begin{equation}
  \pi(\theta, z | y) = \frac{\pi(y | z, \theta) \pi(z | \theta) \pi(\theta)}{\int_{\class{T}} \pi(y | z, \vartheta) \pi(z | \vartheta) \pi(\vartheta) \dd{\vartheta} \meas{M}(\dd{z})},
\end{equation}
where the normalizing constant in the denominator is typically intractable. Since global dependence is typically induced by $\theta$, we will focus on the problem of simulating according to the conditional $\pi(\theta | y, z)$, and write $x = \braces{y, z}$. MCMC algorithms circumvent the intractable normalizing constant by simulating a Markov chain with stationary distribution $\pi(\theta | x)$. A generic construction of such a Markov chain is the \emph{Metropolis-Hastings algorithm} \citep{metropolis1953equation, hastings1970monte}, which from a given state $\theta$ of the Markov chain proposes a move to a value $\vartheta$ according to the transition kernel $\kappa(\vartheta | \theta)$ (w.r.t. Lebesgue measure), and accepts that move with probability
\begin{equation}
  \alpha_{\mathrm{MH}}(\theta, \vartheta) = \op{min}{1, \frac{\kappa(\theta | \vartheta)}{\kappa(\vartheta | \theta)} \frac{\pi(\vartheta, x)}{\pi(\theta, x)}}.
\end{equation}
Critically, the \emph{posterior odds} $\pi(\vartheta | x) / \pi(\theta | x) = \pi(\vartheta, x) / \pi(\theta, x)$ may be evaluated without knowledge of the normalizing constant. Should the move be rejected, the Markov chain remains at $\theta$. If the transition kernel $\kappa$ satisfies some basic requirements, that acceptance procedure gives rise to a Markov chain with stationary distribution corresponding to the posterior. The acceptance decision is usually implemented by drawing $u \sim \op{Unif}{0, 1}$, evaluating $\alpha_{\mathrm{MH}}(\theta, \vartheta)$, and accepting if $u < \alpha_{\mathrm{MH}}(\theta, \vartheta)$. We refer the reader to \cite{tierney1994markov} for a derivation of Metropolis-Hastings in general state spaces with non-Lebesgue dominating measure.

Since the accept-reject procedure involves the evaluation of the posterior odds, it cannot be directly applied whenever $\pi(\theta, x)$ is not available in closed form. It is useful to distinguish two such settings. In the \emph{intractable posterior} setting, the joint density $\pi(\theta, x)$ itself cannot be evaluated, e.g.\ when the likelihood is defined as an intractable integral over additional nuisance (latent) variables. In the \emph{doubly intractable} setting, the likelihood is known only up to a parameter-dependent normalizing constant $Z(\theta)$, i.e.\ $\pi(x | \theta) = \tilde{\pi}(x | \theta) / Z(\theta)$ with $Z(\theta)$ intractable; here the posterior odds retain the factor $Z(\theta) / Z(\vartheta)$, which does not cancel and cannot be evaluated. The latter terminology reflects that a second intractable normalizing constant compounds the usual intractable normalizing constant of the posterior. In both settings, however, the obstruction is the same -- the unnormalised posterior $\pi(\theta, x)$ cannot be evaluated pointwise -- and the solution proposed in this paper addresses them uniformly. We therefore adopt \emph{intractable posterior} as a blanket term for both throughout the paper, reserving the qualifier \emph{doubly intractable} for when the specific source of intractability is relevant.

Various approaches have been developed to address intractable likelihoods. For low-dimensional integrals, quadrature approximations can be computed to very low bias at moderate cost. The conventional approach in higher dimensions is to enlarge the state space with additional auxiliary variables, chosen so that the intractable terms either become amenable to unbiased estimation or cancel in the acceptance ratio. The extended posterior may then be targeted by \emph{augmented Gibbs samplers} \citep{tanner1987calculation} or \emph{pseudo-marginal algorithms} \citep{andrieu2009pseudo}, where the former is usually understood as updating additional nuisance variables according to their full conditional, whereas the latter updates $\theta$ and nuisance variables jointly, and may be seen as carrying out unbiased estimation of the marginal likelihood. The main drawback of such algorithms is that they often result in slower mixing compared to the corresponding \emph{marginal} algorithm on the smaller state space. While some counterexamples are known in Gibbs samplers \citep{liu1999parameter}, it is more typical for augmentation to penalise mixing \citep{liu1995covariance, roberts1997updating}. For integrals over real vectors, constant-time mixing can often be achieved by choosing an appropriate parameterisation of the model \citep{papaspiliopoulos2020scalable, papaspiliopoulos2023scalable}, in which case augmented Gibbs samplers are usually optimal. However, outside of ordinary Lebesgue integrals, such a parameterisation is not always apparent or available. As for pseudo-marginal algorithms, they always mix more slowly than corresponding marginal algorithms \citep{andrieu2015convergence}, and their mixing degrades proportionally to the variance of the likelihood estimator \citep{sherlock2015efficiency}, which is asymptotically $O(e^n)$ for IID data. Studies of optimal scaling for pseudo-marginal algorithms usually recommend holding the estimator variance constant \citep{sherlock2015efficiency, doucet2015efficient}, which necessitates $O(n)$ sampling effort, and results in $O(n^{2})$ cost. Absent such stabilisation effort, mixing time inflates exponentially in $n$ \citep{doucet2015efficient}. A related class of algorithms, sometimes referred to as \emph{ratio-pseudo-marginal} or \emph{coupled MCMC} uses an unbiased estimate of the posterior odds, rather than estimating the numerator and denominator terms individually \citep{nicholls2012coupled}. The ratio estimator benefits from posterior concentration, which can bound the variance in $n$ under IID data. Conversely, with the exception of the \emph{exchange algorithm} \citep{murray2006mcmc} for doubly intractable likelihoods, ratio-pseudo-marginal algorithms are not necessarily \emph{exact}, where we understand exactness as the error being entirely due to Monte Carlo approximation \citep{beskos2006exact}.

With that in mind, we seek to implement the exact marginal algorithm through the 2-coin \emph{Bernoulli factory} approach, which we introduce below before discussing its limitations in the large $n$ regime --- an exponential slowdown analogous to the scaling of mixing time in pseudo-marginal algorithms. We address these limitations through a divide-and-conquer composition of ordinary 2-coin algorithms. Our approach preserves exactness on the marginal chain even when the corresponding pseudo-marginal algorithm does not. We apply divide-and-conquer at the Bernoulli factory level, as opposed to the more common notion of breaking the data into shards, carrying out separate inferences for each shard and unifying those \citep{neiswanger2013asymptotically, dai2023bayesian}, with \emph{Bayesian fusion} even preserving our notion of exactness. While these methods were developed with parallelization and privacy in mind, they could in principle sidestep the exponential scaling by breaking the data into $\Theta(n)$ shards of constant size each. Critically, since they operate on batches of samples from each sub-posterior, they do not naturally accommodate models that require conditioning on $z$ to factorise, e.g.\ stochastic volatility models. Moreover, to our knowledge, there are no theoretical guarantees for the scaling of Bayesian fusion as the number of sub-posteriors increases. Therefore, we deem it preferable not to distribute the inference across multiple chains, unless required for parallel computation or privacy. Furthermore, our solution allows for within-chain parallel computation under some assumptions on the dependence structure of $z$.

Our methodology yields a marginal-MCMC implementation at polynomial cost. In models that naturally factorise as data accrues, it can be exploited to yield improvements in scaling degree over pseudo-marginal algorithms. We require that for given $\theta$, a bounded unbiased estimator is available for at least one of the following quantities: the likelihood itself, its reciprocal, or the likelihood ratio. This compares favourably to pseudo-marginal algorithms, which require likelihood estimators, or the exchange algorithm, which only applies to doubly intractable models. Pseudo-marginal algorithms do not absolutely require a bounded estimator, but their theoretical properties when using unbounded estimators are known to be poor \citep{andrieu2009pseudo}. The complexity of our marginal implementation depends on which of the estimators is available.

The paper proceeds with an introduction to the basic Bernoulli factory MCMC algorithm in Section~\ref{sec:2c}, before discussing its limitations in the large data regime in Section~\ref{sec:scaleprob}. Section~\ref{sec:dnc} describes a divide-and-conquer-style recursive composition of the 2-coin algorithm, which we label the \emph{Cascading 2-coin algorithm} (C2C). This is followed by a complexity analysis in the $n$-limit in Section~\ref{sec:complexity}. We close with computer experiments on a diffusion model and a Cox process in Section~\ref{sec:diff} and Section~\ref{sec:cox} respectively, where we empirically verify the benefits of the C2C.

\subsection{The 2-coin Bernoulli Factory for Barker's Acceptance Probability}
\label{sec:2c}

The basic insight of \cite{gonccalves2023exact} is that explicit knowledge of the acceptance probability is not required for the operation of the MCMC algorithm; in fact, it is sufficient to evaluate the event $\braces{u < \alpha_{\mathrm{MH}}(\theta, \vartheta)}$, or an event of equal probability. This is an instance of the \emph{Bernoulli factory problem}, which consists of simulating coins with probability of heads $h(p)$ from a coin with probability of heads equal to $p$, where $p$ may be unknown --- see e.g.\ \cite{nacu2005fast} for a reference on the problem. In the simplest instance, the input coins follow from a factorisation of form
\begin{equation}
  \label{eq:likfac}
  \kappa(\vartheta | \theta) \pi(\theta, x) = c(\vartheta,\theta) p(\theta),
\end{equation}
where $c(\vartheta, \theta)$ is a tractable upper bound on the left-hand side, and $p(\theta) \in [0, 1]$ is unknown but simulable. An event of probability $p(\theta)$ is typically obtained from a bounded unbiased estimator $\hat{p}(\theta) \in [0, 1]$ (a.s.) and the event $u < \hat{p}(\theta)$; such an estimator is available whenever the problem admits a pseudo-marginal solution with a bounded likelihood estimator. Note that the variance of $\hat{p}(\theta)$ is of no consequence. Thus we have expressed $\alpha_{\mathrm{MH}}(\theta, \vartheta)$ as a function of intractable probabilities, and given an appropriate Bernoulli factory, we could generate an $\alpha_{\mathrm{MH}}(\theta, \vartheta)$-coin from the $p(\theta)$- and $p(\vartheta)$-coins.

\begin{example}[Intractable mixture]\label{ex:run}
  Let $x$ have intractable mixture density $\pi(x | \theta) = \int \pi(x | \theta, \psi) \pi(\psi | \theta) \dd{\psi}$ with $\sup_{\psi} \pi(x | \theta, \psi) < \infty$ for given $x, \theta$. Set $p(\theta) = \pi(x | \theta) / \sup_{\psi} \pi(x | \theta, \psi)$, and observe that
  \begin{equation}
    p(\theta) = \op{E}[\psi]{\pi(x | \theta, \psi)} / \sup_{\psi} \pi(x | \theta, \psi),
  \end{equation}
  where the integrand is in $[0, 1]$. Hence, $p(\theta)$-coins are obtained by simulating the event $\braces{u < \pi(x | \theta, \psi) / \sup_{\psi} \pi(x | \theta, \psi)}$.
\end{example}

While no efficient Bernoulli factory is known for the Metropolis-Hastings acceptance decision \citep{latuszynski2013clts}, there are alternative acceptance probabilities which also result in a Markov chain with the correct stationary distribution. In particular, accepting with probability
\begin{equation}
  \alpha_{\mathrm{B}}(\theta, \vartheta) = \frac{\kappa(\theta | \vartheta) \pi(\vartheta, x)}{\kappa(\theta | \vartheta) \pi(\vartheta, x) + \kappa(\vartheta | \theta) \pi(\theta, x)} \leq \alpha_{\mathrm{MH}}(\theta, \vartheta)
\end{equation}
gives rise to \emph{Barker's algorithm} \citep{barker1965monte}. Due to the so-called \emph{Peskun ordering} \citep{peskun1973optimum, tierney1994markov}, the Metropolis-Hastings algorithm is more efficient than Barker's algorithm when it can be implemented at the same cost. However, the asymptotic variance of Barker's algorithm when estimating the expectation of test functions is roughly no worse than twice the variance of the Metropolis-Hastings algorithm \citep{latuszynski2013clts}. Accordingly, in instances where the Metropolis-Hastings algorithm is more difficult to implement, Barker's algorithm may be preferred. Indeed, writing the Barker acceptance decision in terms of the factorization \eqref{eq:likfac} gives
\begin{equation}
  \alpha_{\mathrm{B}}(\theta, \vartheta) = \frac{c(\theta, \vartheta) p(\vartheta)}{c(\theta, \vartheta) p(\vartheta) + c(\vartheta, \theta) p(\theta)},
\end{equation}
so that the corresponding Bernoulli factory problem is addressed by the \emph{2-coin algorithm} \citep{gonccalves2023exact}, which operates as shown in Figure~\ref{fig:twocoin} with $c_{1} = c(\theta, \vartheta)$, $p_{1} = p(\vartheta)$, $c_{2} = c(\vartheta, \theta)$, $p_{2} = p(\theta)$. Glossing over the tractable factors, the algorithm accounts for the relative magnitude of $p(\theta)$ and $p(\vartheta)$ by repeatedly flipping coins with either probability, until the first coin to come up heads wins the race. Applying that Bernoulli factory, we obtain the Barker acceptance decision without direct reference to the acceptance probability, which implements the marginal algorithm. This paradigm has been applied to Bayesian inference for Wright-Fisher diffusions \citep{gonccalves2017barker}, jump-diffusion models \citep{gonccalves2023exact}, diffusion-driven Cox processes \citep{gonccalves2023bexact}, models with intractable priors \citep{liechty2009shadow, vats2022efficient}, and intractable proposals \citep{kakkad2024exact}. Alternatively, the \emph{flipped} 2-coin algorithm \citep{vats2022efficient} simulates an $\alpha_{\mathrm{B}}(\theta, \vartheta)$-coin by using an upper bound $\tilde{c}(\vartheta, \theta)$ on $\kappa(\vartheta | \theta)^{-1} \pi(\theta, x)^{-1}$, such that $\kappa(\vartheta | \theta)^{-1} \pi(\theta, x)^{-1} = \tilde{c}(\vartheta, \theta) \tilde{p}(\theta)$ and $\tilde{p}(\theta) \in [0, 1]$. Therefore, we can implement a 2-coin algorithm in scenarios where there is no upper bound on the likelihood, or when it is more natural to estimate the inverse likelihood, the latter of which is not an option in the pseudo-marginal approach. Indeed, any factorization of $\alpha_{\mathrm{B}}$ which satisfies the 2-coin representation results in a valid 2-coin algorithm.

\begin{figure}[ht]
  \centering
  \begin{tikzpicture}[edge from parent/.style={draw,latex-}, level 1/.style={level distance=15mm}, level 2/.style={level distance=15mm, sibling distance=80mm}]
  \node (start) at (0, 0) {\textbf{start}}; 
  \node (f0) at (0, -1) {$C_{0}$};
  \draw[->] (start) -- (f0);
  \node (f1) at (-2, -2) {$C_{1}$};
  \draw[->, bend right] (f0) to node[midway, above left] {$\frac{c_{1}}{c_{1} + c_{2}}$} (f1);
  \draw[->, bend right] (f1) to node[very near start, below right] {$1 - p_{1}$} (f0);
  \node (f2) at (2, -2) {$C_{2}$};
  \draw[->, bend left] (f0) to node[midway, above right] {$\frac{c_{2}}{c_{1} + c_{2}}$} (f2);
  \draw[->, bend left] (f2) to node[very near start, below left] {$1 - p_{2}$} (f0);
  \node (r1) at (-4.5, -2) {\textbf{return} 1};
  \draw[->] (f1) -- node[midway, below] {$p_{1}$} (r1);
  \node (r0) at (4.5, -2) {\textbf{return} 0};
  \draw[->] (f2) -- node[midway, below] {$p_{2}$} (r0);
  \end{tikzpicture}
  \caption{Probability flow diagram of the 2-coin algorithm, generating a compound coin with probability $c_{1} p_{1} / \braces{c_{1} p_{1} + c_{2} p_{2}}$ based on elementary coins of probability $p_{1}$ and $p_{2}$. Nodes ($C_{0}$, $C_{1}$, $C_{2}$) refer to coin flips; edges give the probabilities of moving to the corresponding node. The algorithm's expected number of loops (ENL) corresponds to the expected number of visits to $C_{0}$.}
  \label{fig:twocoin}
\end{figure}

Compared to augmented algorithms, Barker's 2-coin algorithm has the distinct advantage of implementing the marginal algorithm and preserving its mixing properties. That being said, we show in Section~\ref{sec:scaleprob} that the 2-coin algorithm can have complexity as bad as $n e^{\Omega(n)}$ for a data size $n$.

\subsection{The Bernoulli Factory Scaling Problem}
\label{sec:scaleprob}

Consider the data growth regime where the likelihood admits the factorization $\pi(x | \theta) = \prod_{i=1}^{n} c_{i}(\theta) p_{i}(\theta)$ and $p_{i}(\theta)$ is a probability according to which we can simulate coins independently. A fundamental issue with the vanilla 2-coin algorithm of Section~\ref{sec:2c} is that the aggregate bound $c(\theta) = \prod_{i=1}^{n} c_{i}(\theta)$ slackens as data accumulates, so that $p(\theta) = \prod_{i=1}^{n} p_{i}(\theta)$ shrinks. The reason is structural: a $p_{i}(\theta)$-coin is simulated by sampling an unbiased estimator $\hat{p}_{i}(\theta)$ of the probability $p_{i}(\theta)$ with $0 \le \hat{p}_{i}(\theta) \le 1$ almost surely, returning heads when $u < \hat{p}_{i}(\theta)$, which fires with probability $p_{i}(\theta)$. For the bound to hold almost surely, $c_{i}(\theta)$ must exceed the largest value of $c_{i}(\theta) p_{i}(\theta)$, hence strictly exceed its mean, so $p_{i}(\theta) < 1$; tightening it would require evaluating the intractable likelihood itself. The slack is thus unavoidable and, being multiplicative, compounds: if $p_{i}(\theta) \le \delta < 1$ then $p(\theta) \le \delta^{n}$. For example, with exchangeable data and constant slack $p_{i}(\theta) \equiv \rho$ one gets $p(\theta) = \rho^{n}$, already of order $10^{-5}$ for the benign values $\rho = 0.9$, $n = 100$; the same compounding occurs whenever each $c_{i}(\theta)$ over-covers its factor by a non-vanishing margin, regardless of the model or estimator.

To quantify this effect in a way that does not depend on the per-step cost, we study the expected number of loops. The algorithm of Figure~\ref{fig:twocoin} repeats until it returns a decision, so its cost is the number of passes through the starting node times the per-pass cost; the latter is at most $O(n)$, but the number of passes is geometric with termination probability governed by $p(\theta)$ and $p(\vartheta)$, and it is here that the exponential blow-up resides. The expected number of loops therefore isolates the scaling pathology from implementation detail and admits a closed form, which motivates the following definition.

\begin{definition}[Expected number of loops]
  \label{def:enl}
  The \emph{expected number of loops} (ENL) of an algorithm is the average number of times the algorithm visits the starting node ($C_{0}$ in Figure~\ref{fig:twocoin}) before terminating.
\end{definition}

In particular, the number of loops in the vanilla 2-coin algorithm has distribution $\op{Geom}{(c(\theta, \vartheta) p(\vartheta) + c(\vartheta, \theta) p(\theta)) / (c(\theta, \vartheta) + c(\vartheta, \theta))}$, with expectation
\begin{equation}
  \label{eq:enl2c}
  \frac{c(\theta, \vartheta) + c(\vartheta, \theta)}{(c(\theta, \vartheta) p(\vartheta) + c(\vartheta, \theta) p(\theta))} \geq \frac{1}{\op{max}[a = \theta, \vartheta] \braces{\prod_{i=1}^{n} p_{i}(a)}}.
\end{equation}
Since $p_{i}(\theta)$ is smaller than 1 in the non-trivial case where $c_{i}(\theta)$ is loose, the lower bound on the ENL increases exponentially with $n$. Flipping the second coin has cost as high as $n$, therefore resulting in $n e^{\Omega(n)}$ complexity. Notice the analogy to pseudo-marginal algorithms with single-sample likelihood estimator, where iteration time remains linear but mixing time is exponential in $n$, resulting in the same overall scaling. Unlike pseudo-marginal, the cost is driven by the looseness of the upper bound, rather than the variance of the estimator.

\begin{remark}[Lower bound on 2-coin ENL]
  A simple finite-horizon argument makes the exponential growth explicit without requiring the per-factor probabilities to be bounded away from one uniformly in $n$ --- a property we cannot guarantee, since for some factors the bound $c_{i}$ may tighten as $n$ grows. Fix any horizon $n_{0} \in \mathbb{N}$. In the non-trivial regime where the bounds are loose, $p_{i}(\theta) < 1$ for each $i$, so $\delta \coloneqq \max_{1 \le i \le n_{0}} p_{i}(\theta) < 1$ as a maximum of finitely many terms strictly below one. Then $\prod_{i=1}^{n} p_{i}(\theta) \le \delta^{n}$ for every $n \le n_{0}$, and the ENL lower bound $\parens*{\prod_{i=1}^{n} p_{i}}^{-1}$ in \eqref{eq:enl2c} grows at least as fast as $\delta^{-n}$ on this horizon. As $n_{0}$ is arbitrary, this establishes exponential cost for every fixed data size encountered in practice. A rigorous asymptotic statement, under mild conditions, is provided in Supplement~\ref{sec:lemmata}: Propositions \ref{prop:expE} and \ref{prop:expP} establish (i) an exponential lower bound on the expected ENL and (ii) an exponentially small probability that the ENL grows only sub-exponentially.
\end{remark}
 
Various improvements can be made within the 2-coin framework. The \emph{Portkey 2-coin algorithm} returns 0 with a fixed probability at every loop of the 2-coin algorithm, while still preserving the correct stationary distribution \citep{vats2022efficient}. This allows for early stopping of a 2-coin simulation, which is very helpful e.g.\ when proposing a value $\vartheta$ for which the 2-coin ENL is high and posterior probability is low. Nonetheless, the Portkey algorithm does not fundamentally address the scaling problem, since the rejection rate will merely increase as data accumulates. Moreover, the Barker acceptance probability allows for arbitrary refactorization to yield a different 2-coin algorithm. In particular,
\begin{equation}
    \alpha_{\mathrm{B}}(\theta, \vartheta) 
    = \frac{\frac{\kappa(\theta | \vartheta)}{\kappa(\vartheta | \theta)} \frac{\pi(\vartheta, x)}{\pi(\theta, x)}}{1 + \frac{\kappa(\theta | \vartheta)}{\kappa(\vartheta | \theta)} \frac{\pi(\vartheta, x)}{\pi(\theta, x)}} 
    = \frac{1}{\frac{\kappa(\vartheta | \theta)}{\kappa(\theta | \vartheta)} \frac{\pi(\theta, x)}{\pi(\vartheta, x)} + 1},
\end{equation}
so if there is a factorization of the odds $\kappa(\theta | \vartheta) \pi(\vartheta, x) / \braces{\kappa(\vartheta | \theta)\pi(\theta, x)}$ or its reciprocal into $d(\theta, \vartheta) q(\theta, \vartheta)$ such that $d(\theta, \vartheta)$ is tractable and bounded and $q(\theta, \vartheta)$ is a probability according to which we can simulate coin flips, then we obtain the alternative 2-coin representation
\begin{equation}\label{eq:rtfrm}
  \alpha_{\mathrm{B}}(\theta, \vartheta) = \frac{d(\theta, \vartheta) q(\theta, \vartheta)}{d(\theta, \vartheta) q(\theta, \vartheta) + 1} \quad \text{or} \quad \frac{1}{1+d(\vartheta, \theta) q(\vartheta, \theta)}.
\end{equation}
Such a factorization is exploited in \cite{stumpf2025exact}. The main benefit of this 2-coin algorithm lies in the option of exploiting posterior concentration, which usually results in a contraction of the step size at the same rate \citep{schmon2022optimal}. This can yield a lower bound on $q(\theta, \vartheta)$ that only decays at rate $e^{\sqrt{n}}$. We elaborate on the construction and properties of such coins in Section~\ref{sec:leafcost}. Moreover, if we deflate the step size at the faster rate $1/n$, we incur a mixing penalty of $O(n)$ but tighten the bound to $O(1)$, yielding a quadratic algorithm. Our goal, then, is to improve on that degree, and extend the class of models for which we achieve polynomial complexity.

We proceed from the diagnosis that the aggregation of $p_{i}(\theta)$-coins to a $p(\theta)$-coin \emph{within} a single 2-coin algorithm is the root of the scaling problem, since such aggregation leads to an exponential decay of the second coin. Instead, we follow the divide-and-conquer paradigm and obtain coins that correspond to Barker acceptance coins for subsets of the data. This method shall allow us to trade effort that scales exponentially in data size for effort that only scales polynomially, analogous to variance stabilization in pseudo-marginal algorithms.

\section{Recursive 2-coin Algorithms}
\label{sec:dnc}

We proceed with specifying the most general setting where our solution is applicable.

\begin{assumption}[Factor decomposition]
  \label{ass:dec}
  For some notion of data size $n$, the Barker acceptance odds factorise as $\alpha_{\mathrm{B}}(\theta, \vartheta) / \alpha_{\mathrm{B}}(\vartheta, \theta) = \prod_{i=1}^{n} f_{i}(\theta, \vartheta)$, and each factor can be represented as 
  \begin{equation}
    f_{i}(\theta, \vartheta) = \frac{d_{i}(\theta, \vartheta) q_{i}(\theta, \vartheta)}{d_{i}(\vartheta, \theta) q_{i}(\vartheta, \theta)},
  \end{equation}
  where $d_{i}(\theta, \vartheta) > 0$ is tractable and $q_{i}(\theta, \vartheta)$ is a probability according to which we can simulate coins. Finally, let the $q_{i}(\theta, \vartheta)$-coins be independent of each other, conditional on $x$.
\end{assumption}

Such a factorization arises naturally through (conditional) independence, such as in IID models and Markov processes, e.g.\ the ones we examine in Sections \ref{sec:diff} and \ref{sec:cox}. In the elementary case of Section~\ref{sec:scaleprob}, the identity is simply $d_{i}(\vartheta, \theta) q_{i}(\vartheta, \theta) = c_{i}(\theta) p_{i}(\theta)$. We choose the odds formulation due to its greater generality and the possibility of aggregating odds by simple multiplication. Probabilities follow from odds through a 1-to-1 transformation.

Our strategy is to exploit the factor decomposition obtained through Assumption~\ref{ass:dec} by simulating cheap elementary coins with odds equal to the factors, and aggregating those to the more expensive coin with odds equal to the product. As a starting point, let $\alpha_{\mathrm{B}}(\theta, \vartheta) / \alpha_{\mathrm{B}}(\vartheta, \theta) = h_{0}(\theta, \vartheta)$, and suppose that for all $\theta, \vartheta \in \class{T}$, we can decompose $h_{0}(\theta, \vartheta) = \prod_{j=1}^{m} h_{j}(\theta, \vartheta)$. We can think of each $h_{j}$ as a subset of the $\braces{f_{i}}$ factors. Though we present the methods of this section within the context of Barker updates to parameter vectors, they can in principle be applied in any context where the acceptance odds factorise, e.g.\ to Barker updates of $z$. Nonetheless, since global dependence is typically induced by $\theta$, we focus on the parameter update setting.

The key observation is that the 2-coin algorithm can itself be fed by 2-coin outputs. If the inputs correspond to the Barker probability for subsets of the data, the output aggregates those to the Barker probability for the joint data. Define the probabilities $r_{j}(\theta, \vartheta)$ with odds $r_{j}(\theta, \vartheta) / r_{j}(\vartheta, \theta) = h_{j}(\theta, \vartheta)$, notice that $r_{0}(\theta, \vartheta) = \alpha_{\mathrm{B}}(\theta, \vartheta)$, and consider the odds of a 2-coin algorithm with (potentially intractable) parameters $p_{1} = \prod_{j} r_{j}(\theta, \vartheta)$, $p_{2} = \prod_{j} r_{j}(\vartheta, \theta)$ and $c_{1} = c_{2}$. The odds of the output are
\begin{equation}
  \frac{\prod_{j=1}^{m} r_{j}(\theta, \vartheta)}{\prod_{j=1}^{m} r_{j}(\vartheta, \theta)} = \prod_{j=1}^{m} h_{j}(\theta, \vartheta) = h_{0}(\theta, \vartheta) = \frac{r_{0}(\theta, \vartheta)}{r_{0}(\vartheta, \theta)},
\end{equation}
so the probability of 1 is precisely $\alpha_{\mathrm{B}}(\theta, \vartheta)$. It directly follows that the cost of this merger is exponential in $m$, with ENL
\begin{equation}
\label{eq:enl2c_b}
  \frac{1}{\prod_{j=1}^{m} r_{j}(\theta, \vartheta) + \prod_{j=1}^{m} r_{j}(\vartheta, \theta)} = \frac{r_{0}(\theta, \vartheta)}{\prod_{j=1}^{m} r_{j}(\theta, \vartheta)} = \frac{r_{0}(\vartheta, \theta)}{\prod_{j=1}^{m} r_{j}(\vartheta, \theta)}.
\end{equation}
Nonetheless, this perspective suggests that we can replace a single, slow 2-coin simulation that targets $\alpha_{\mathrm{B}}(\theta, \vartheta)$ with a tree-shaped recursive application of faster 2-coin simulations, thereby avoiding the exponential scaling. We shall derive the basic algorithm in Section~\ref{sec:c2c} and discuss extensions in Section~\ref{sec:portkey}.

\subsection{The Cascading 2-coin Algorithm}
\label{sec:c2c}

We now formalisea recursive partition of the factors and devise a recursive aggregation scheme for intractable factors. Let $\#_{k}: \braces{1,\dots,n} \to \braces{1, 2}^{k}$ satisfy $\#_{k}(i) = (\#_{k-1}(i),1)$ or $\#_{k}(i) = (\#_{k-1}(i),2)$, which defines a binary tree. Thus, for $j_{1} \dots j_{k} \in \braces{1, 2}^{k}$ and $h_{j_{1} \dots j_{k}} = \prod_{i: \#_{k}(i) = j_{1} \dots j_{k}} f_{i}$,
\begin{equation}
  h_{j_{1} \dots j_{k}} = h_{j_{1} \dots j_{k},1} \times h_{j_{1} \dots j_{k},2}.
\end{equation}
Accordingly, as we move from the leaves towards the root of the tree, we accumulate more and more factors. Moreover, setting $r_{j_{1} \dots j_{k}}(\theta, \vartheta) / r_{j_{1} \dots j_{k}}(\vartheta, \theta) = h_{j_{1} \dots j_{k}}(\theta, \vartheta)$, we obtain a coin of probability $r_{j_{1} \dots j_{k}}(\theta, \vartheta)$ from applying 2-coin to coins of probability $r_{j_{1} \dots j_{k},1}(\theta, \vartheta)$ and $r_{j_{1} \dots j_{k},2}(\theta, \vartheta)$: flip these two coins to obtain $C_{1}$ and $C_{2}$, return their common value when $C_{1} = C_{2}$, and repeat otherwise. This is exactly a 2-coin algorithm, in which $C_{1}$ is the selection coin, choosing side $1$ with probability $r_{j_{1} \dots j_{k},1}(\theta, \vartheta)$ and side $2$ with probability $r_{j_{1} \dots j_{k},1}(\vartheta, \theta) = 1 - r_{j_{1} \dots j_{k},1}(\theta, \vartheta)$, while $C_{2}$ provides the acceptance coin, of probability $r_{j_{1} \dots j_{k},2}(\theta, \vartheta)$ on side $1$ (accepted when $C_{2} = 1$) and $r_{j_{1} \dots j_{k},2}(\vartheta, \theta)$ on side $2$ (accepted when $C_{2} = 0$). Consequently the step returns a coin of probability
\begin{equation}
  \frac{r_{j_{1} \dots j_{k},1}(\theta, \vartheta)\, r_{j_{1} \dots j_{k},2}(\theta, \vartheta)}
       {r_{j_{1} \dots j_{k},1}(\theta, \vartheta)\, r_{j_{1} \dots j_{k},2}(\theta, \vartheta)
        + r_{j_{1} \dots j_{k},1}(\vartheta, \theta)\, r_{j_{1} \dots j_{k},2}(\vartheta, \theta)}
  = r_{j_{1} \dots j_{k}}(\theta, \vartheta).
  \label{eq:merge}
\end{equation}
On that basis, the \emph{Cascading 2-coin algorithm} (C2C) applies this merge recursively from the leaves to the root, each input coin being itself produced by a recursive C2C call on the corresponding child subtree, with the base case at the leaf level $\ell$ handled by a conventional 2-coin algorithm call. Applying the output probability of the 2-coin algorithm recursively, it follows that this procedure results in a coin of probability $r_{0}(\theta, \vartheta) = \alpha_{\mathrm{B}}(\theta, \vartheta)$ at the root. We provide a recursive pseudocode specification below.

\begin{algorithm}
  \begin{algorithmic}
    \Function{C2C}{$\class{F}_{j_{1}\dots j_{k}} = \braces{f_{i} : \#_{k}(i) = j_{1}\dots j_{k}}$}
    \If{$k = \ell$}
    \Return \Call{2C}{$\class{F}_{j_{1}\dots j_{k}}$}
    \Else
    \Loop
    \State $C_{1} \gets$ \Call{C2C}{$\class{F}_{j_{1}\dots j_{k},1}$}
    \State $C_{2} \gets$ \Call{C2C}{$\class{F}_{j_{1}\dots j_{k},2}$}
    \If{$C_{1} = C_{2}$}
    \Return $C_{1}$
    \EndIf
    \EndLoop
    \EndIf
    \EndFunction
  \end{algorithmic}
  \caption{Defining $\class{F}_{j_{1}\dots j_{k}} = \braces{f_{i} : \#_{k}(i) = j_{1}\dots j_{k}}$, \texttt{C2C}($\class{F}_{j_{1}\dots j_{k}}$) yields coin flips with probability $r_{j_{1}\dots j_{k}}(\theta, \vartheta)$. The call \texttt{2C}($\class{F}_{j_{1}\dots j_{k}}$) is to an implementation of the 2-coin algorithm that yields coin flips with probability $r_{j_{1}\dots j_{k}}(\theta, \vartheta)$ independently.}
  \label{alg:c2c}
\end{algorithm}

 \begin{proposition}[Cascading 2-coin algorithm]\label{prop:c2c}
   Under Assumption~\ref{ass:dec} (independence of coins), the marginal output probability of Algorithm~\ref{alg:c2c} is $r_{0}(\theta, \vartheta)$.
 \end{proposition}

We further visualisethe dynamics of the algorithm for $\ell = 2$ in Figure~\ref{fig:c2c}, where the factor batches are fed into 2-coin algorithms at the leaves. Using this schematic for orientation, it is instructive to consider the sequence of mergers towards the root. For instance, once the coins at $r_{11}$ and $r_{12}$ have reached agreement and merged to a coin flip at $r_{1}$, they are not affected by any disagreement at $r_{21}$ and $r_{22}$ - the flip at $r_{1}$ is locked in until the flip at $r_{2}$ is independently generated. Only in case of disagreement between $r_{1}$ and $r_{2}$ does the algorithm need to return to the leaves below $r_{1}$. This contrasts with the vanilla 2-coin algorithm, where all $n$ coin flips must be 1 simultaneously. As a side benefit, since longer computations are more amenable to parallel computation than shorter ones, the C2C is more naturally parallelized through a recursive \emph{fork-join} pattern. For example, we can split computation across the nodes $r_{1}$ and $r_{2}$, which have to communicate relatively rarely, whereas in vanilla 2-coin, computations have to be unified every time the second coin is composed from its sub-coins.

\begin{figure}
  \centering
  \begin{tikzpicture}[edge from parent/.style={draw,latex-}, level 1/.style={level distance=15mm}, level 2/.style={level distance=15mm, sibling distance=80mm}, level 3/.style={level distance=15mm, sibling distance=40mm}]
  \node {\textbf{output}}
    child {node[circle,draw] {$r_{0}$}
      child {node[circle,draw] {$r_{1}$}
        child {node[circle,draw] {$r_{11}$} child {node {$h_{11}$}} edge from parent node[above left] {$\rho_{11}$}}
        child {node[circle,draw] {$r_{12}$} child {node {$h_{12}$}} edge from parent node[above right] {$\rho_{12}$}}
        edge from parent node[above left] {$\rho_{1} = \tau_{1} (\rho_{11} + \rho_{12})$} 
      }
      child {node[circle,draw] {$r_{2}$}
        child {node[circle,draw] {$r_{21}$} child {node {$h_{21}$}} edge from parent node[above left] {$\rho_{21}$}}
        child {node[circle,draw] {$r_{22}$} child {node {$h_{22}$}} edge from parent node[above right] {$\rho_{22}$}}
        edge from parent node[above right] {$\rho_{2} = \tau_{2} (\rho_{21} + \rho_{22})$} 
      }
    edge from parent node[left] {$\rho_{0} = \tau_{0} (\rho_{1} + \rho_{2})$}
    };
  \end{tikzpicture}
  \caption{Schematic of the C2C for $\ell = 2$. Nodes are 2-coin simulations and their label corresponds to their probability of generating 1. Links are labelled according to the expected cost of generating an output when moving up the link. The elementary factors are fed in at the bottom of the tree.}
  \label{fig:c2c}
\end{figure}

In spite of the C2C's simple recursive definition and the intuition for its computational benefits that we have acquired, rigorous runtime analysis calls for a more careful exposition. As Algorithm~\ref{alg:c2c} shows, calls to \textsc{C2C()} are agnostic about the specifics of the 2-coin algorithm at the leaves, and as seen in Section~\ref{sec:scaleprob}, this 2-coin algorithm can have various runtime properties, depending on the exact factorization and model. For the cost of the C2C itself, all that matters is the ENL at each node to achieve a merger and propagate a coin toss towards the root. We shall label this quantity $\tau_{j_{1} \dots j_{k}}$ at node $j_{1} \dots j_{k}$, while the total cost of producing a coin toss at that node is $\rho_{j_{1} \dots j_{k}}$. In particular, $\rho_{j_{1} \dots j_{\ell}}$ is the cost of a 2-coin simulation at the leaves. As in \eqref{eq:enl2c_b}, $\tau_{j_{1} \dots j_{k}}$ has representation
\begin{equation}
  \tau_{j_{1} \dots j_{k}} = \frac{r_{j_{1} \dots j_{k}}(\theta, \vartheta)}{r_{j_{1} \dots j_{k},1}(\theta, \vartheta) r_{j_{1} \dots j_{k},2}(\theta, \vartheta)}. \qquad (k < \ell)
\end{equation}
From these elements, we obtain a bound on the expected cost of the whole procedure.

\begin{proposition}[Cost of C2C]
  \label{prop:c2ccost}
  Let $\rho_{0}$ be the expected cost of obtaining a $r_{0}(\theta, \vartheta)$-coin toss with the C2C. It has representation
  \begin{equation}
    \rho_{0} = \sum_{j_{1} \dots j_{\ell} \in \braces{1, 2}^{\ell}} \rho_{j_{1} \dots j_{\ell}} \tau_{0} \prod_{k=1}^{\ell - 1} \tau_{j_{1} \dots j_{k}},
  \end{equation}
  where we use $\tau_{0}$ to denote the ENL at the root.
\end{proposition}

In the equation above, each summand corresponds to the path to one of the leaves at $j_{1} \dots j_{\ell}$, while the product $\tau_{0} \prod_{k=1}^{\ell - 1} \tau_{j_{1} \dots j_{k}}$ therein is the expected number of times that the 2-coin algorithm at that leaf is executed. Figure~\ref{fig:c2c} illustrates the accumulation of computational cost towards the root. Our analysis in Section~\ref{sec:complexity} bears out that the recursion indeed avoids exponential scaling.

\subsection{Recursive Portkey Barker Algorithm}
\label{sec:portkey}

Given the existence of disadvantageous data splits where the runtime of the C2C is large, it is useful to introduce the option of aborting the C2C, such that we can make a new attempt with a different partition, or a new proposal. The best starting point to implement such an option is the \emph{Portkey Barker algorithm} \citep{vats2022efficient}. The key observation is that the algorithm which accepts a move $\theta \to \vartheta$ with the probability
\begin{equation}
  \frac{\kappa(\theta | \vartheta) \pi(\vartheta, x)}{b(\theta, \vartheta) + \kappa(\theta | \vartheta) \pi(\vartheta, x) + \kappa(\vartheta | \theta) \pi(\theta, x)},
\end{equation}
is reversible if $b(\theta, \vartheta) = b(\vartheta, \theta)$, i.e.\ we may inflate the probability of rejection as long as we do so in a symmetrical way. While this reduces the efficiency of the algorithm in the Peskun sense, it also gives us the option of terminating the Bernoulli factory early, thereby truncating a simulation with large average runtime. The Portkey 2-coin algorithm returns tails at every iteration with some fixed probability, and decreases the acceptance probability symmetrically, thereby preserving reversibility. We propose a similar extension to the recursive merge algorithm, and demonstrate that it preserves detailed balance.

\begin{proposition}[Cascading portkey 2-coin algorithm]
\label{prop:portkeydnc}
  Consider the Bernoulli factory that aborts with probability $1 - \varpi_{j_{1} \dots j_{k}}$ each time it requires a coin toss at node $j_{1} \dots j_{k}$, instantly rejecting the proposal value. This Bernoulli factory returns heads with probability
  \begin{equation}
    \frac{\kappa(\theta | \vartheta) \pi(\vartheta, x)}{b_{0}(\theta, \vartheta) + \kappa(\theta | \vartheta) \pi(\vartheta, x) + \kappa(\vartheta | \theta) \pi(\theta, x)}
  \end{equation}
  for some non-negative function $b_{0}(\cdot, \cdot)$ such that $b_{0}(\theta, \vartheta) = b_{0}(\vartheta, \theta)$, so it implements a reversible acceptance decision.
\end{proposition}

This extension of the Portkey feature presents us with two options for improving the C2C. Firstly, within the C2C, we can apply the Portkey 2-coin algorithm at the leaves, e.g.\ mitigating long runtimes caused by bad proposals. Secondly, we can bias all merge nodes in the C2C towards rejection, thereby accelerating termination in the instance of unbalanced data partitions for which the unbiased C2C is slow. As for setting the escape probabilities, the framework allows for setting node-specific values $\varpi_{j_{1} \dots j_{k}}$. In particular, we may choose not to escape at all on certain levels. In fact, since a slow node towards the root causes many coin flips at its descendant nodes, we deem it sufficient to set a uniform non-zero escape probability at the leaf level $\ell$, i.e.\ $\varpi_{j_{1} \dots j_{\ell}} = \varpi > 0$. As for scaling with $n$, and assuming that the expected number of 2-coin loops is linear in data size, we should be willing to expend at least linear effort on average before escaping, which implies that $1 - \varpi$ should scale as $\Theta(1/n)$.

Notice that different partitions of the product terms yield distinct versions of the Portkey Barker algorithm. However, the strategy of selecting a partition randomly at each iteration remains valid. This is because the resulting chain's transition distribution is a finite mixture of individual transition distributions, all sharing the same invariant distribution. Consequently, the invariant distribution is preserved in the overall chain.

\section{Complexity Analysis}
\label{sec:complexity}

For a full complexity analysis of the C2C, it is useful to examine the merge cost in the recursion separately from the cost of the 2-coin algorithm at the leaves. This is due to them being partially driven by different properties of the model, the emphasis of this paper on the properties of the recursion, and the fact that the runtime at the leaves can be characterized in isolation.

Starting with Proposition~\ref{prop:c2ccost}, we presume equal data splits such that leaf batch size is $\Theta(2^{-\ell} n)$, and loosely posit the upper bound $\varrho(n, \ell) \geq \rho_{j_{1} \dots j_{\ell}}$. This leaves us with the \emph{merge cost}
\begin{equation}
  \omega = \tau_{0} \sum_{j_{1} \dots j_{\ell} \in \braces{1, 2}^{\ell}} \prod_{k=1}^{\ell - 1} \tau_{j_{1} \dots j_{k}},
\end{equation}
which is the expected number of 2-coin outputs required for one iteration of the full C2C and, therefore, $\rho_{0} \leq \omega \varrho(n, \ell)$. This naturally raises the question of how to minimise $\rho_{0}$ as a function of $\ell$. In fact, $\omega$ necessarily grows with $\ell$, but $\varrho(n, \ell)$ shrinks with $\ell$, inducing a trade-off. Assuming that $\omega$ is indeed polynomial, this trade-off is best addressed by eliminating exponential terms in $\varrho(n, \ell)$. For example, in the scenario of Section~\ref{sec:scaleprob} with independent data and likelihood estimators, the scaling is $\varrho(n, \ell) = O(2^{-\ell} n e^{2^{-\ell} n})$, so setting $2^{\ell} \propto n$ yields $\varrho(n, \ell) = O(1)$. Under this choice, the behaviour of $\omega$ in the large-$n$ limit becomes the key factor in the scaling of the overall algorithm. 


The analysis of the merge cost is presented in Section~\ref{sec:mergecost} below. This is followed by a more precise analysis of the leaf cost in Section~\ref{sec:leafcost}, resulting in a characterization of the models for which we achieve different polynomial degrees in Section~\ref{sec:synth}.

\subsection{Merge Cost Analysis}
\label{sec:mergecost}

The most immediate way of studying the scaling of the merge cost results from setting the MCMC step size to 0, i.e.\ $\vartheta = \theta$, or from copying the same data across all leaves. As we establish later on in Theorem~\ref{thm:gibbsscale}, the resulting scaling corresponds to an average-case analysis for non-0 step sizes or heterogeneous data. 

\begin{proposition}[Scaling of balanced merge cost]
  \label{prop:balscale}
  Set $\vartheta = \theta$. Then, $\omega = 4^{\ell}$. Alternatively, if all $h_{j_{1} \dots j_{\ell}}$ are identical, then $\omega \leq 4^{\ell}$.
\end{proposition}

Note that if at every merge step we split the data into $m$ batches, the corresponding expression is $\omega = m^{\ell_{m}} 2^{\ell_{m} (m - 1)}$. For instance, if we set $m = 2^{k}$ and use $\ell_{m} = \ell/k$ to hold the batch size at the leaves constant, the merge cost is $2^{\ell + (2^{k} - 1) \ell / k}$, which increases in $k \in \mathbb{N}$, and suggests that the 2-way split is optimal. Based on that result, we proceed with analysis of $\omega$ when $\vartheta \neq \theta$, with an eye on recovering the same scaling. The following proposition characterizes $\omega$ in the most general setting.

\begin{proposition}[Merge cost with randomized assignment]
  \label{prop:morand}
  Under the setting of Assumption~\ref{ass:dec}, let $\varsigma$ be a random permutation of $\braces{1, \dots, n}$, such that $\braces{f_{\varsigma(1)}, \dots, f_{\varsigma(n)}}$ is exchangeable. Moreover, suppose that partitions are applied to the shuffled set $\braces{f_{\varsigma(1)}, \dots, f_{\varsigma(n)}}$, such that the batch assignment is uniformly random. Finally, let the data be split equally at each node of the tree, such that the batch size at $j_{1}\dots j_{k}$ is $\Theta(2^{-k} n)$. Then,
  \begin{equation}
    \op{E}[\varsigma]{\omega}
    = 2^{\ell} \parens*{1 + 2 r_{0}(\vartheta, \theta) \sum_{j=1}^{2^{\ell}-1} \op{E}[\varsigma]{\prod_{i=1}^{nj 2^{-\ell}}f_{\varsigma(i)}(\theta, \vartheta)}}.
  \end{equation}
  Note that we immediately recover $\op{E}[\varsigma]{\omega} = 4^{\ell}$ for $\theta = \vartheta$.
\end{proposition}

The basic insight is that control of $\omega$ in $n$ depends on preventing the divergence of the posterior odds of the various subsamples. If the odds remain stable as $n$ increases, there are $2^{\ell}$ stable terms on the RHS, which with the leading factor $2^{\ell}$ results in a RHS of order $4^{\ell}$. In controlling the expected likelihood ratio, we must account for two countervailing forces as $n \to \infty$. On one hand, multiplying further factors could result in divergence of the ratio. On the other hand, if the model is smooth, the usual $n^{-1/2}$ scaling of the proposal reduces the magnitude of each ratio, and in sufficiently regular models, posterior concentration pulls the MCMC algorithm towards a region in $\class{T}$ where gradients are small. The coupling of the asymptotics of the algorithm with the asymptotics of the data generating process (DGP), which we deem necessary to achieve a well-behaved limit, makes for a fairly intricate setting. Therefore, we will operate within the class of single-parameter IID models, though the basic insights transfer to more complicated models, as long as they are sufficiently smooth and the posterior concentrates. 

To carry out the analysis where posterior concentration is endogenous, we must also be precise about the context under which the C2C is applied. We consider a correctly specified 2-component Gibbs sampler in stationarity, where the full conditional update to $\pi(\theta | x_{1:n})$ is carried out by Barker/C2C-within-Gibbs. We may then study the computational cost of that update as $n$ gets large. As to the model and the DGP, we impose the following assumptions.

\begin{assumption}[DGP]
  \label{ass:dgp}
  Let $y_{1:n} = \braces{y_{1} \dots y_{n}}$ be distributed independently and identically according to the DGP $\meas{P}_{0}$ with induced density $\pi(y_{1:n} | \theta_{0}) = \prod_{i=1}^{n} \pi(y_{i} | \theta_{0})$, for some true parameter value $\theta_{0}$. The correctly specified model has latent variable representation $\pi(y_{i} | \theta) = \int \pi(y_{i}, z_{i} | \theta) \dd{z_{i}}$.
\end{assumption}

\begin{assumption}[Smoothness]
  \label{ass:smooth}
   Let $\class{B}$ be a neighbourhood of $\theta_{0}$. On that neighbourhood, $\pi(y_{i} | \theta)$ satisfies that
  \begin{itemize}
      \item $\theta \mapsto \log \pi(y_{i} | \theta)$ is twice continuously differentiable $\meas{P}_{0}$-a.s.
      \item the score identity holds, i.e.\ $\op{E}[\meas{P}_{0}]{\partial_{\theta_{0}} \log \pi(y_{i} | \theta_{0})} = 0$.
      \item the derivatives have MGF bounds
      \begin{equation}
          \sup_{\theta \in \class{B}} \op{E}[\meas{P}_{0}]{\exp \braces{\lambda 
          \abs{\partial_{\theta} \log \pi(y_{i} | \theta)}}} < \infty, \quad \op{E}[\meas{P}_{0}]{\exp \braces{\lambda \sup_{\theta \in \class{B}} \abs{\partial_{\theta}^{2} \log \pi(y_{i} | \theta)}}} < \infty,
      \end{equation}
      for small enough $\lambda$.
  \end{itemize}
\end{assumption}

\begin{assumption}[Posterior and proposal concentration]
  \label{ass:post}
  The correctly specified $n$-posterior $\pi(\theta_{n} | y_{1:n})$ has a Bernstein-von-Mises type limit such that it converges in total variation to $\op{N}{\theta_{n}; \theta_{0}, I_{0}^{-1} / n}$ with Fisher information $I_{0}$. Moreover, $\pi(\theta_{n} | x_{1:n})$ concentrates at the same rate. The proposal follows $\vartheta_{n} | \theta_{n} \sim \op{Unif}{\theta_{n} \pm \delta / \sqrt{n}}$ for some step size $\delta$, and in the $n$-limit achieves constant acceptance rate in the $\pi(\theta | x_{1:n})$ update.
\end{assumption}

Assumption~\ref{ass:smooth} is a fairly standard set of conditions in Bayesian asymptotics and high-dimensional statistics that can be leveraged to bound likelihood ratios by way of Taylor expansions. Assumption~\ref{ass:post} is satisfied by a broad range of parametric models, and the posterior concentration rate of $n^{-1/2}$ is also achieved in some dependent processes such as Markov models \citep{borwanker1971bernstein}. Under these assumptions, we obtain our main result.

\begin{theorem}[Merge cost scaling]
  \label{thm:gibbsscale}
  Consider a Gibbs sampler at stationarity with random inputs $y_{1:n}$, and let $\meas{G}_{n}$ be the measure with law
  \begin{equation}
    \pi(y_{1:n} | \theta_{0}) \pi(z_{1:n}, \theta_{n} | y_{1:n}) \op{Unif}{\vartheta_{n}; \theta_{n} \pm \delta / \sqrt{n}}.
  \end{equation}
  We presume that $\ell$ is scaled with $n$ such that $n \to \infty \implies \ell \to \infty$. Under Assumptions~\ref{ass:dgp},~\ref{ass:smooth} and~\ref{ass:post}, the $n$-sample merge cost $\omega_{n}$ satisfies
  \begin{equation}
    4^{-\ell} \omega_{n} = O_{\meas{G}_{n}}(1).
  \end{equation}
\end{theorem}

This is also known as \emph{stochastic boundedness}, in that it ensures that none of the probability mass runs away to infinity as $n$ increases. We note that stochastic boundedness does not exclude aberrant behaviour far out in the tails of the posterior. As a consequence, absent stronger regularity conditions, $\omega_{n}$ is not necessarily integrable with respect to $\meas{G}_{n}$. That said, this applies regardless of $n$ and is therefore orthogonal to our emphases on the large $n$ regime, and the result guarantees that any given MCMC run is unlikely to be problematic. Moreover, we can mitigate such issues by using the Portkey C2C introduced above. Alternatively, integrability for such algorithms is often established by restricting $\theta$ to a compact set.

We have also derived variants of this theorem for more general data dependence structures, which require additional regularity conditions on the latent model and the nature of the dependence. Since they work by exploiting the same essential features, i.e.\ posterior concentration and smoothness, we prefer the conditionally IID setting for the sake of a clearer presentation. Generally speaking, we expect the result to hold for models where observables and latent variables satisfy a notion of sufficiently fast mixing, e.g.\ Markov processes and Gaussian processes with sufficiently fast covariance function decay. Indeed, our computer experiments in Section~\ref{sec:exp} empirically verify that the scaling holds for such models.

\subsection{Leaf Cost Analysis}
\label{sec:leafcost}

As for the analysis of the cost $\varrho(n, \ell)$ at the leaves, we distinguish the \emph{leaf iteration cost} $\gamma_{j_{1} \dots j_{\ell}}$ of running a single loop of the 2-coin algorithm on a batch of $\Theta(2^{-\ell}n)$ factors, and the \emph{leaf ENL} $\tau_{j_{1} \dots j_{\ell}}$ of the same algorithm.

We lead with leaf ENL since it makes the exponential contribution. As discussed in Section~\ref{sec:scaleprob}, in the case where the 2-coin algorithm is obtained from a bounded likelihood estimator, the leaf ENL on a batch of $2^{-\ell} n$ factors is typically $\Theta(e^{2^{-\ell} n})$. We can mitigate this by constructing coins that derive from a likelihood ratio estimator, and exploit continuity in $\theta$ in conjunction with posterior concentration. For the purpose of this discussion, we label the availability of ratio-derived coins the \emph{ratio estimator property}. While the construction of likelihood estimators usually follows trivially from the development of the model, it bears elaborating on the generic constructions of coins from ratio estimators, and the circumstances where they apply.

\begin{example}[Perfect Sampling Estimators]
\label{ex:perfect}
  Suppose that $f_{i}(\theta, \vartheta) = \pi(x_{i} | \vartheta) / \pi(x_{i} | \theta)$ and there is a tractable augmentation scheme such that $\pi(x_{i} | \theta) = \int \pi(x_{i}, \psi | \theta) \dd{\psi}$. Then,
  \begin{equation}
    f_{i}(\theta, \vartheta) = \int \frac{\pi(x_{i}, \psi | \vartheta)}{\pi(x_{i}, \psi | \theta)} \pi(\psi | x_{i}, \theta) \dd{\psi},
  \end{equation}
  i.e.\ there is an unbiased estimator for $f_{i}(\theta, \vartheta)$ if we can implement perfect sampling from $\pi(\psi | x_{i}, \theta)$, e.g.\ by rejection sampling or \emph{coupling from the past}. For instance, such algorithms exist for some population genetics models \citep{fearnhead2001perfect}. We may then factorise
  \begin{equation}
    f_{i}(\theta, \vartheta) = d_{i}(\theta, \vartheta) q_{i}(\theta, \vartheta), \quad q_{i}(\theta, \vartheta) = \frac{\frac{\pi(x_{i} | \vartheta)}{\pi(x_{i} | \theta)}}{\sup_{\psi} \frac{\pi(x_{i}, \psi | \vartheta)}{\pi(x_{i}, \psi | \theta)}} = \parens*{\sup_{\psi} \frac{\pi(\psi | x_{i}, \vartheta)}{\pi(\psi | x_{i}, \theta)}}^{-1}.
  \end{equation}
  A noteworthy special case arises when working with an unnormalised law $\tilde{\pi}(x_{i} | \theta)$ from which we can sample perfectly. Then,
  \begin{equation}
    f_{i}(\theta, \vartheta) = \frac{\tilde{\pi}(x_{i} | \vartheta)}{\tilde{\pi}(x_{i} | \theta)} \int \frac{\tilde{\pi}(x' | \theta)}{\tilde{\pi}(x\ | \vartheta)} \pi(x' | \vartheta) \dd{x'},
  \end{equation}
  which is exploited e.g.\ in \cite{murray2006mcmc}.
\end{example}

\begin{example}[Poisson Estimators]
\label{ex:poissonest}
  Let $l_{i}(\theta, \vartheta) = \log f_{i}(\theta, \vartheta)$ have unbiased estimator $\hat{l}_{i}(\theta, \vartheta; \psi)$ with random element $\psi$, bounded within $l_{i}^{\downarrow}(\theta, \vartheta) = \inf_{\psi} \hat{l}_{i}(\theta, \vartheta; \psi)$ and $l_{i}^{\uparrow}(\theta, \vartheta) = \sup_{\psi} \hat{l}_{i}(\theta, \vartheta; \psi)$. Then, for $M \sim \op{Poisson}{l_{i}^{\uparrow}(\theta, \vartheta) - l_{i}^{\downarrow}(\theta, \vartheta)}$, $u_{1}, \dots, u_{M} \sim \op{Unif}{l_{i}^{\downarrow}(\theta, \vartheta), l_{i}^{\uparrow}(\theta, \vartheta)}$,
  \begin{equation}
    f_{i}(\theta, \vartheta) = e^{l_{i}^{\uparrow}(\theta, \vartheta)} \op{E}[M, \psi_{1}, \dots, \psi_{M}, u_{1}, \dots, u_{M}]{\prod_{m = 1}^{M} \op{1}{\hat{l}_{i}(\theta, \vartheta; \psi_{m}) > u_{m}}}.
  \end{equation}
  Note that evaluation of this estimator can be terminated as soon as one factor is 0. We can exploit the estimator to construct a 2-coin algorithm where we factorise
  \begin{equation}
    f_{i}(\theta, \vartheta) = d_{i}(\theta, \vartheta) q_{i}(\theta, \vartheta), \quad q_{i}(\theta, \vartheta) = e^{l_{i}(\theta, \vartheta) - l_{i}^{\uparrow}(\theta, \vartheta)}.
  \end{equation}
  Appropriate log-likelihood estimators are often available when the likelihood is implied by a Girsanov-type change of measure, e.g.\ in continuous time and space models such as diffusions and Poisson processes, and we exploit them in Sections \ref{sec:diff} and \ref{sec:cox}. The Poisson estimator is generalized by the Russian roulette estimation framework \citep{lyne2015russian}, which can in principle replace the exponential link function between the unbiased estimator and the likelihood ratio by other functions.
\end{example}

We proceed with analysis of the leaf ENL with $q_{i}$-coins. The basic insight is that if we have a Lipschitz-type bound on $f_{i}(\theta, \vartheta)$ and the optimal proposal scales as $|\theta - \vartheta| \leq \delta / \sqrt{n}$ due to posterior concentration, then at leaf $j_{1} \dots j_{\ell}$ the second coin has probability $\prod_{\braces{i: \#_{\ell}(i) = j_{1} \dots j_{\ell}}} q_{i}(\theta, \vartheta) \geq e^{-(K/\sqrt{n})|\theta - \vartheta|}$, and the leaf ENL is $e^{O(2^{-\ell} \sqrt{n})}$. The following proposition formalizes this notion without requiring a uniform Lipschitz constant.

\begin{proposition}[Leaf ENL scaling]
\label{prop:leafenl}
  In the setting of Assumptions~\ref{ass:dgp} and~\ref{ass:post}, let $f_{i}(\theta, \vartheta) = d_{i}(\theta, \vartheta) q_{i}(\theta, \vartheta)$, with $q_{i}$ constructed as in Example~\ref{ex:perfect} and $\meas{G}_{n}$ defined as in Theorem~\ref{thm:gibbsscale}. Moreover, let $\pi(\psi | x_{i}, \theta)$ have MGF bound
  \begin{equation}
    \op{E}[y_{1} \sim \pi(y_{1} | \theta_{0}), z_{1} \sim \pi(z_{1} | y_{1}, \theta_{n})]{\exp\braces*{\lambda \sup_{\theta \in \class{B}, \psi} |\partial_{\theta} \log \pi(\psi | x_{1}, \theta)|}}< \infty
  \end{equation}
  for small enough $\lambda > 0$ and $\theta_{n} \in \class{B}$. Then, $\tau_{j_{1} \dots j_{\ell}}$ satisfies
  \begin{equation}
    \frac{2^{\ell}}{\sqrt{n}} \log \tau_{j_{1} \dots j_{\ell}} = O_{\meas{G}_{n}}(1).
  \end{equation}
\end{proposition}

An analogous result applies to the construction in Example~\ref{ex:poissonest}. The upshot is that if the ratio estimation property applies and the posterior concentrates, setting $4^{\ell} \propto n$ eliminates the exponential in the leaf ENL, and accordingly $\omega = O(n)$. If it does not apply, setting $2^{\ell} \propto n$ eliminates the exponential, and accordingly $\omega = O(n^{2})$.

We now consider the smaller matter of leaf iteration cost $\gamma_{j_{1} \dots j_{\ell}}$, determined by the cost of flipping the second coin in the 2-coin algorithm. For models where the likelihood ratio factorises through independence, $\gamma_{j_{1} \dots j_{\ell}}$ usually corresponds to the batch size $\Theta(2^{-\ell} n)$, since one estimator has to be evaluated for each data unit at the leaf. If instead it factorises through annealing, we expect that $\gamma_{j_{1} \dots j_{\ell}}$ is at least $O(n)$, irrespective of batch size, as each coin involves the entire data set. Some settings allow for further savings. For instance, note that the Poisson estimator in Example~\ref{ex:poissonest} consists of up to $M$ evaluations of $\hat{l}_{i}$, which is $\sup_{\psi} \hat{l}_{i}(\theta, \vartheta; \psi) - \inf_{\psi} \hat{l}_{i}(\theta, \vartheta; \psi)$ on average. This rate is subject to contraction at rate $1 / \sqrt{n}$ under posterior concentration. Eventually the cost of flipping the $q_{i}$-coin is dominated by the constant cost of generating $M$ at a very small rate. We can avoid that cost by generating the index of the next factor where $M > 0$, which is $\Theta(\sqrt{n})$, eliminating the constant cost. Thus, under this setting, flipping $q_{i}$-coins has cost $O(1/\sqrt{n})$, which adds up to a leaf iteration cost of $O(1)$ over a batch size of $\Theta(\sqrt{n})$. We call this the \emph{batching property}, and it is generally implied by Poisson-type ratio estimators.

\subsection{Synthesis} 
\label{sec:synth}

In the light of this analysis, the key insight is that the C2C can mitigate a bound on the acceptance odds of order $e^{O(n^p)}$ by setting $2^{\ell} \propto n^{p}$ and paying merge cost $O(n^{2p})$. This keeps the bound constant at the leaves, thereby eliminating the exponential scaling of leaf ENL and achieving polynomial cost.

The specific outcome follows from whether the leaf 2-coin is implemented through an (inverse) likelihood estimator or likelihood ratio estimator. We summarise this in the following taxonomy, subject to regularity conditions such as posterior concentration through e.g.\ Assumption~\ref{ass:smooth}.
\begin{itemize}
  \item If the ratio estimation property does not apply, we set $2^{\ell} \propto n$, so leaf ENL is $O(1)$ and merge cost is $O(n^{2})$. Leaf iteration cost is $O(1)$ due to constant batch size, and overall cost is $O(n^{2})$.
  \item If the ratio estimation property applies, we set $4^{\ell} \propto n$, so leaf ENL is $O(1)$ and merge cost is $O(n)$. For models without the batching property, leaf iteration cost is $O(\sqrt{n})$ and overall complexity is $O(n^{3/2})$. If the batching property does apply, leaf iteration cost is $O(1)$ and overall complexity is $O(n)$.
\end{itemize}

It follows that in the first instance, we recover the analogy with correctly tuned pseudo-marginal algorithms, which are quadratic as well. In the second instance, it is superior to pseudo-marginal algorithms, and it is here that the C2C delivers the greatest benefits. Such scenarios often exhibit a Girsanov-type intractable likelihood. Conversely, in doubly intractable models, the exchange algorithm usually delivers $O(n)$ scaling, so we expect advantages over exchange at most expect in ill-behaved, pre-asymptotic regimes, which benefit from a strictly marginal algorithm.

\section{Computer Experiments}
\label{sec:exp}

The algorithm is broadly applicable, but its greatest gains arise in models defined through an intractable Girsanov-type change of measure, which are pervasive in continuous-time and continuous-space formulations. Diffusions and Cox processes are the most prominent examples: both are highly flexible classes that can be extended almost arbitrarily \citep{beskos2006exact, gonccalves2018exact, gonccalves2023exact, gonccalves2023bexact, stumpf2025exact}, with some of these works already relying on the 2-coin or C2C algorithm. For all of these models, we contend that the C2C is the state-of-the-art solution for exact Bayesian MCMC inference. Accordingly, in this section, we illustrate the performance of C2C on diffusions (Section~\ref{sec:diff}) and Cox processes (Section~\ref{sec:cox}), using both simulated and real data. In each case we compare C2C against the relevant existing state of the art --- the vanilla 2-coin algorithm for the diffusion models, and augmented Gibbs sampling for the Cox process. 




\subsection{Application to Diffusion Models}
\label{sec:diff}

Diffusion-driven models are the class in which the proposed algorithm has the greatest immediate impact. The intractable likelihood of a complete diffusion path has a particular form that is amenable to an efficient unbiased ratio estimator via the Poisson estimator of Example~\ref{ex:poissonest}. Diffusion processes are typically specified through a \emph{stochastic differential equation} (SDE); for simplicity of presentation we restrict ourselves to the unit-volatility class
\begin{equation}
  \dd{X_{t}} = \beta_{\theta}(X_{t}) \dd{t} + \dd{W_{t}}.
\end{equation}
Under appropriate regularity conditions on $\beta_{\theta}$, this SDE defines a Markovian diffusion process, and we write $B_{\theta}(u) = \int^{u} \beta_{\theta}(a) \dd{a}$. Processes with non-constant diffusion coefficient are implicitly covered, since the Lamperti transform maps them to a process with constant coefficient. The corresponding discretely-observed likelihood is intractable except for a few special cases, such as the Ornstein-Uhlenbeck process with $\beta_{\theta}(X_{t}) = -\theta X_{t}$. A key ingredient in exact inference is the complete-data likelihood that includes the diffusion bridges. For discrete observations $x_{t_{0}}, \dots, x_{t_{n}}$ (taking the role of $y$ in our general notation) and imputed diffusion bridges $x_{(t_{0}, t_{1})}, \dots, x_{(t_{n-1}, t_{n})}$ (taking the role of $z$), the complete-data likelihood (with respect to Wiener measure) is
\begin{equation}
  \pi(x_{(0, t_{n}]} | \theta, x_{0}) = \prod_{i=1}^{n} \op{exp}{B_{\theta}(x_{t_{i}}) - B_{\theta}(x_{t_{i-1}}) - \int_{t_{i-1}}^{t_{i}} \varphi_{\theta}(x_{s}) \dd{s}}, \quad \varphi_{\theta} = (\beta_{\theta}^{2} + \beta_{\theta}') / 2
\end{equation}
and we notice that the time integral over $\varphi_{\theta}$ is intractable. 

Exact inference for diffusion models, in the sense of being subject only to Monte Carlo error, is a challenging and long-standing inference problem, and relies on algorithms that simulate diffusion paths from their exact probability law by retrospective rejection sampling \citep{beskos2008factorisation,beskos2006exact,beskos2009monte,sermaidis2013markov}, the so-called Exact Algorithms (EA). Various classes of diffusion processes are addressed through different versions of the algorithm, but many important diffusion processes, such as Wright-Fisher diffusions, fall outside their domain of validity, as their density with respect to Brownian bridge measure is not lower bounded. Notably, looser conditions apply within the 2-coin Barker MCMC algorithm framework, allowing inference e.g.\ for Wright-Fisher diffusions \citep{gonccalves2023exact, gonccalves2017barker}. Where previously the vanilla 2-coin algorithm with its scaling weakness was the only option, the C2C allows for considerable performance improvements, lowering complexity from exponential to sub-quadratic in the time horizon. Moreover, even where EA-based MCMC is available, it requires further augmentation of the state space, whereas the Barker MCMC chain consists only of the parameters and the unobserved diffusion path. 

We implement a 2-component Gibbs sampler (Algorithm~\ref{alg:diffusion}) with Barker updates to $\pi(\theta | x_{[0, t_{n}]})$ and $\pi(x_{(t_{i-1}, t_{i-1})} | \theta, x_{t_{i-1}}, x_{t_{i}})$, noting that the diffusion bridge updates factorise at the observation times. We leverage C2C to obtain a scalable update to $\theta$, while our benchmark consists of updating $\theta$ with the vanilla 2-coin algorithm. The diffusion bridge update is carried out as in \citep{gonccalves2017barker}. As discussed in Supplement~\ref{app:diff}, the complete-data likelihood permits the construction of ratio-coins through a Poisson-style estimator, and it therefore has the ratio estimation property.

\begin{algorithm}
  \begin{algorithmic}
    \Loop
    \State $\theta \sim \pi(\theta | x_{[0, t_{n}]})$ \Comment{by way of C2C Barker [Algorithm~\ref{alg:c2c}] with $q_{i}$-factors}
    \For{i = 1\dots n}
    \State $x_{(t_{i-1}, t_{i})} \sim \pi(x_{(t_{i-1}, t_{i-1})} | \theta, x_{t_{i-1}}, x_{t_{i}})$ \Comment{by way of standard 2-coin Barker \cite{gonccalves2017barker}}
    \EndFor
    \EndLoop
  \end{algorithmic}
  \caption{Marginal Barker-within Gibbs sampler for the diffusion model.}
  \label{alg:diffusion}
\end{algorithm}

\subsubsection{Simulation Study}


In this section, we carry out a systematic scalability study of the C2C for Bayesian inference in discretely-observed diffusion processes, comparing it against the vanilla 2-coin algorithm. We test the C2C on the diffusion model with SDE
\begin{equation}
  \dd{X_{t}} = \op{tanh}{\theta - X_{t}} \dd{t} + \dd{W_{t}}
\end{equation}
and prior $\theta \sim \op{N}{0, 1}$, where the goal is to infer $\theta$. Due to the sigmoid shape of the $\tanh$-function, this describes a heavier-tailed process that approaches an Ornstein-Uhlenbeck process for $X_{t} \to \theta$. We simulate observations at times $t_{i} = 0, 1/4, 2/4, \dots, n/4$ according to the tanh-SDE with $\theta = 0$. For increasing values of $n$, we then run the Barker-within-Gibbs algorithm for 10{,}000 iterations. We set the C2C tree depth to $\ell = \log_{4} n$, the $\theta$-update uses a random walk proposal with step size $8 / \sqrt{n}$, and the Portkey setting is $1 - \varpi = 1/n$. The benchmark algorithm uses the same step size, but we apply the Portkey setting $1 - \varpi = e^{-\sqrt{n}}$, which keeps the probability of escaping constant.

\begin{figure}[ht]
  \centering
  \includegraphics[scale=.5]{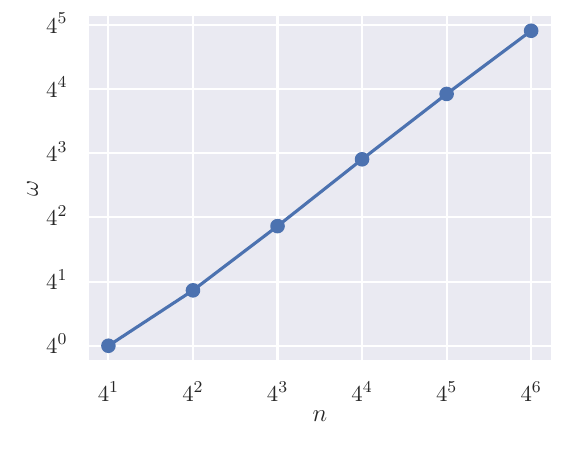}
  \includegraphics[scale=.5]{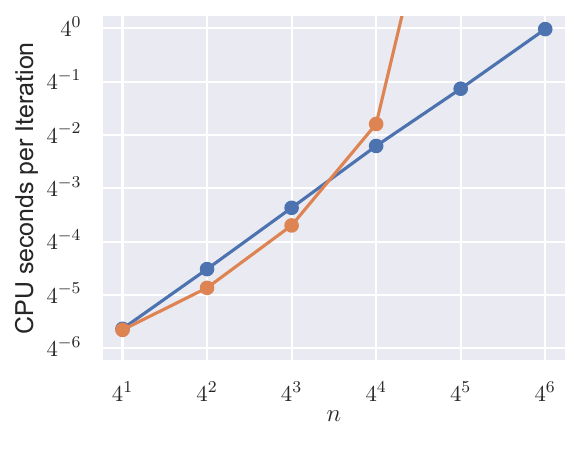}
  \includegraphics[scale=.5]{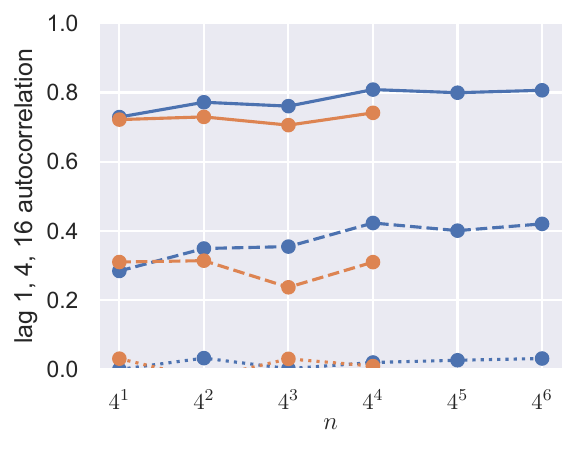}
  \caption{[Diffusion simulation study] Relationship between $n$ and various performance characteristics, averaged over a Markov chain of 10{,}000 iterations, for the algorithm with the C2C (blue) and vanilla 2-coin (orange). Iteration time refers to the $\theta$-update only. Autocorrelation estimates are provided for lag 1 (solid), 4 (dashed), and 16 (dotted). The estimates for vanilla 2-coin with $n=4^5$ and $n=4^6$ are based only on 100 samples due to exponentially increasing computational cost. We therefore omit the autocorrelation estimates for those runs.}
  \label{fig:tanhscale}
\end{figure}

Performance characteristics of the $\theta$-update are displayed in Figure~\ref{fig:tanhscale}, which we summarise as follows:
\begin{itemize}
  \item The mixing of the algorithm, as assessed by lag 1 autocorrelation, remains stable as $n$ increases, yielding a similar and non-decreasing rate of effective samples per iteration for all runs.
  \item In accordance with Theorem~\ref{thm:gibbsscale}, the merge cost $\omega$ is linear in $n$, averaged over the chain.
  \item CPU time per iteration appears to be approximately $O(n^{1.2})$, in accordance with our prediction of $O(n^{1.5})$ when the ratio estimation property applies.
\end{itemize}
Note that while the batching property could be applied to lower complexity to $O(n)$, this requires a tight bound on the ratio-coins, which is not always cheaply evaluated in more complicated diffusion models. We therefore deem this non-optimized version more representative.


\subsubsection{Real data analysis: implied equity correlation}

We model the Cboe 3-month implied-correlation index $X_{t} \in (0,1)$  --- the option-implied average correlation among the S\&P~500 constituents%
--- as a Wright-Fisher diffusion,
\begin{equation}\label{eq:wf}
  \dd{X_{t}} = -\tfrac{\theta_{1}}{2}\,(X_{t}-\theta_{2}) \dd{t} + \sigma\sqrt{X_{t}(1-X_{t})}\,\dd{W_{t}},
\end{equation}
with stationary law $\op{Beta}{\theta_{1}\theta_{2}/\sigma^2,\theta_{1}(1-\theta_{2})/\sigma^2}$. As a deterministic function of continuously quoted option prices, $X_{t}$ is a genuine continuous-time process recorded \emph{exactly} at discrete (weekly) times, so no observation-error layer is needed and the discretely-observed diffusion applies directly. We use $T = 257$ weekly observations (2~December 2018 -- 29~October 2023).

The boundary behaviour also governs the computational cost. Under the Lamperti transform the Girsanov potential $\varphi_{\theta}$ is unbounded near the boundaries and varies sharply. The cost of the standard 2-coin algorithm then becomes  prohibitive in most cases while C2C remains feasible for much more extreme trajectories.

We fit the model with a Barker-within-Gibbs sampler that updates three types of block in turn: the diffusion bridge on each inter-observation interval (each a separate block), the drift pair $(\theta_{1}, \theta_{2})$, and $\sigma$; the two parameter blocks use the C2C formulation. We ran the sampler for $10{,}000$ iterations, initialising the three parameters at their method-of-moments estimates. The resulting posterior means (standard deviations) are $\theta_{1} = 0.173\,(0.046)$, $\theta_{2} = 0.375\,(0.036)$ and $\sigma = 0.098\,(0.0045)$.

\begin{figure}[!ht]
  \centering
  \includegraphics[width=0.9\linewidth]{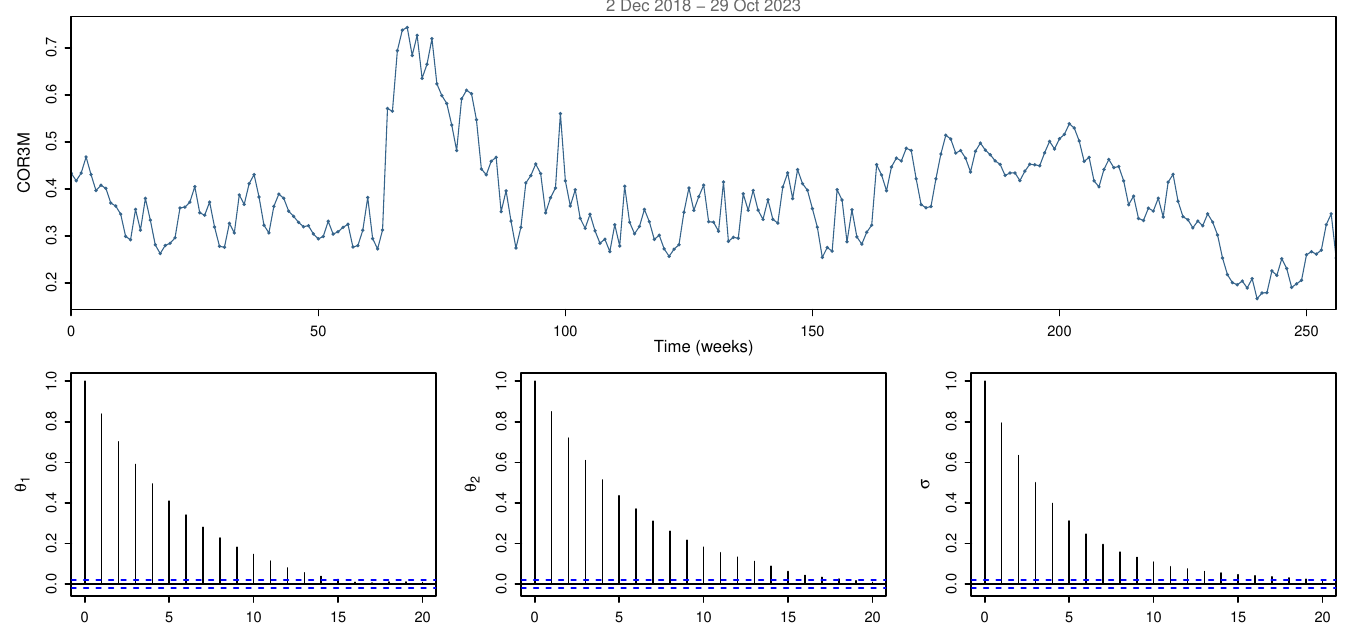}
  \caption{[Diffusion application] Cboe 3-month implied-correlation index (\texttt{COR3M}) and posterior diagnostics. \emph{Top:} the $T=257$ weekly observations. \emph{Bottom:} autocorrelation functions of the posterior draws for
  $\theta_{1}$, $\theta_{2}$ and $\sigma$.}
  \label{fig:cboe}
\end{figure}

The chain mixes well. The uniform random-walk proposals for the two parameter blocks are tuned following optimal-scaling results \citep{agrawal2023optimal}, giving acceptance rates of $0.25$ for $(\theta_{1},\theta_{2})$ and $0.33$ for $\sigma$, with the Portkey truncation triggered in about $6\%$ of iterations for each. The effective sample sizes are $881$, $818$ and $1152$ for $\theta_{1}$, $\theta_{2}$ and $\sigma$ out of the $10{,}000$ draws, and the autocorrelation plots in
Figure~\ref{fig:cboe} show rapid decorrelation. 

\subsection{Application to Cox Processes}
\label{sec:cox}

Having demonstrated the scalability benefits of the C2C in Section~\ref{sec:diff}, we now examine the mixing gains that the C2C enables compared to an augmented chain. We carry out the experiment in a \emph{Cox process} setting, which is usually defined as a Poisson process whose intensity function is random and follows a non-parametric prior. Defining $y$ as a Poisson process on a bounded region $S \subset \mathbb{R}^{d}$, it is often convenient to parameterise the intensity function in terms of a parameter vector $\theta$ and a latent stochastic process $z$ on $S$. Then, $y$ has conditional density function
\begin{equation}\label{eq:pp_like}
  \pi(y | \theta, z) \propto \op{exp}{-\int_{S} \lambda_{\theta}(z_{s}) \dd{s}} \prod_{s \in y} \lambda_{\theta}(z_{s}),
\end{equation}
where we note that the density can be factorised at will over partitions $\braces{T_{1}, \dots, T_{n}}$ of the spatial domain $S$, and the natural data growth regime consists of increasing the volume of $S$. The process $z$ is typically endowed with a (piecewise-)continuous prior, e.g.\ a Gaussian process prior or a diffusion prior, for which the integral is well-defined but intractable. Further augmentation, e.g.\ by way of Poisson thinning, results in a tractable expression, and exact augmented algorithms have been investigated for various priors on $z$ \citep{kottas2007bayesian, adams2009tractable, gonccalves2018exact, gonccalves2023cexact}, most of which accommodate the C2C framework. For Cox-process models, discretization-error-free MCMC is typically implemented with a Gibbs sampler that updates three blocks: $\theta$, $z$, and an auxiliary variable $\psi$ introduced so that the augmented likelihood $\pi(y, \psi | \theta, z)$ is tractable.

While designing a marginal sampler for $z$ is highly model-sensitive, especially with regard to the conditional independence structure of $z$ (e.g.\ Markov), marginal updates to $\pi(\theta | x)$ can usually be implemented and scaled through a C2C Barker update. Indeed, the mixing benefits are usually fully realized by marginalizing the $\theta$-update, even if the $z$-update is not marginalized \citep{van2008partially}. This is known as \emph{Partially collapsed Gibbs} (CGS, Algorithm~\ref{alg:cox}). The algorithm remains valid by resampling the nuisance variable $\psi$ according to $\pi(\psi | \theta, x)$ before updating $z$ according to $\pi(z | \psi, x, \theta)$. The natural benchmark then consists of updating $\theta$ according to the non-marginal $\pi(\theta | z, \psi, y)$, resulting in an ordinary augmented Gibbs sampler (AGS, \citep{gonccalves2023cexact}). We provide further implementation details for our collapsed Gibbs sampler in Supplement~\ref{app:cox}, in particular on construction of $q_{i}$-coins for unbiased estimation of $\pi(y | \theta, z)$, which is achieved through a Poisson-type estimator.

\begin{algorithm}
  \begin{algorithmic}
    \Loop
    \State $\theta \sim \pi(\theta | x)$ \Comment{by way of C2C Barker [Algorithm~\ref{alg:c2c}] with $q_{i}$-factors}
    \State $\psi \sim \pi(\psi | \theta, x)$ \Comment{as in \citep{gonccalves2023cexact}}
    \State $z \sim \pi(z | \theta, \psi, y)$ \Comment{as in \citep{gonccalves2023cexact}}
    \EndLoop
  \end{algorithmic}
  \caption{Marginal Barker within collapsed Gibbs sampler for the Cox process.}
  \label{alg:cox}
\end{algorithm}

\subsubsection{Simulation study}

We test our algorithm on the \emph{level set Cox process} \citep{gonccalves2023cexact}, which has a piecewise-constant intensity function, and where the regions of constant intensity are determined by the level sets of a latent Gaussian process (GP) --- other possibilities include models where the intensity is a continuous function of the GP as in \cite{gonccalves2018exact} and \cite{nolau2026}. In this instance, for some region $S \subset \mathbb{R}^{d}$, the intensity function is
\begin{equation}
  (\lambda_{\theta} \circ z)(s) = \sum_{l=1}^{L} \theta_{l} I_{S_{l}}(s), \quad S_{l} = \braces{s \in S: \upsilon_{l-1} \leq z_{s} < \upsilon_{l}},
\end{equation}
where $I_{S_{l}}$ is the indicator function of $S_{l}$, and the latent function $z$ is drawn from a zero-mean stationary GP. To avoid cubic costs associated with dense GP covariance matrices, $z$ is assumed to be a \emph{nearest neighbour Gaussian process} \citep{datta2016hierarchical}. This has the advantage of scaling linearly with the dimension of the induced multivariate normal distribution. The number of levels $L$ and the GP parameters are fixed. In our experiment, we also fix the thresholds $\braces{\upsilon_{l}: l = 0, 1, \dots, L}$. Thus, the parameter set is $\theta = \braces{\theta_{l}: l = 1, \dots, L}$. 
Details about the C2C and the likelihood estimator are given in Supplement~\ref{app:cox}.


We run both algorithms for 10{,}000 iterations after adaptation on a sequence of squares $S \in \mathbb{R}^{2}$ with common centroid and sides 16, 32, 64 and 128, resulting in areas/data size $4^4$ to $4^7$. Each is vertically split into a low-intensity region with $\theta_{1} = 1/3$ on the left and a high-intensity region $\theta_{2} = 5/3$ on the right. 
Note that the boundary between $S_{1}$ and $S_{2}$ is of length $\sqrt{n}$, which is a computationally favourable case. The resulting integrated intensities over $S$, corresponding to the expected number of observations $|y|$ under the DGP, are $4^4$ to $4^7$, and thus equal to $n$. With both high- and low-intensity regions growing at rate $n$, we expect a posterior concentration rate of $n^{-1/2}$ for both $\theta_{1}$ and $\theta_{2}$. The random walk uniform proposal distributions are tuned to achieve optimal acceptance probabilities, following the theoretical guidelines for Metropolis-Hastings and Barker's algorithm \citep{roberts2001optimal, agrawal2023optimal}, and approximately tracking the posterior concentration rate. We again set the C2C tree depth to $\ell = \log_{4} n$. The Portkey escape probability $1 - \varpi$ in the C2C is set to escape in approximately 10\% of the Barker's iterations.

\begin{figure}[ht]
  \centering
  \includegraphics[scale=.5]{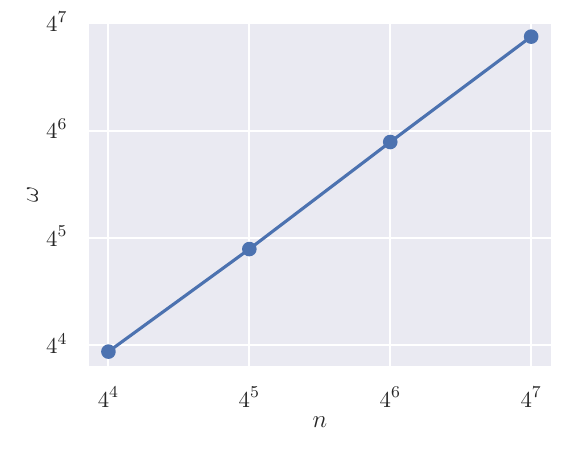}
  \includegraphics[scale=.5]{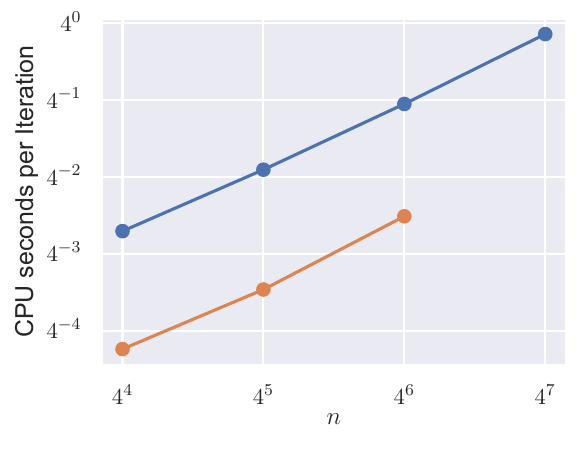}
  \includegraphics[scale=.5]{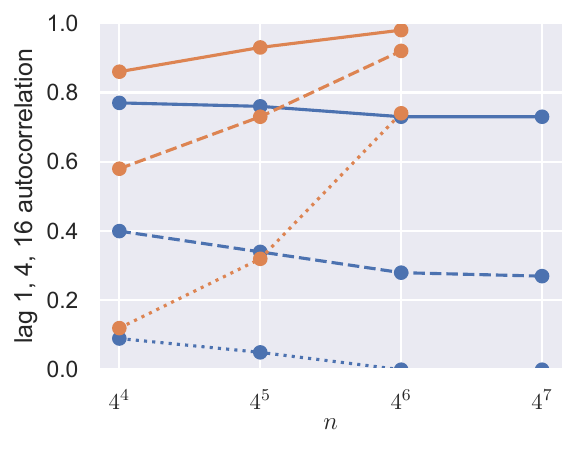}
  \caption{[Cox process simulation study] Relationship between $n$ and various performance characteristics for the $\theta_{2}$-update, averaged over a Markov chain of 10{,}000 iterations, for the CGS (blue) and the AGS (orange). Iteration time refers to the parameter update only. Autocorrelation estimates are provided for lag 1 (solid), 4 (dashed), and 16 (dotted). We omit the AGS run for $n = 4^7$ due to rapidly deteriorating mixing.}
  \label{fig:coxscale}
\end{figure}

We extract three stylised facts from the simulation output shown in Figure~\ref{fig:coxscale}:
\begin{itemize}
  \item The mixing of the partially collapsed Gibbs sampler remains stable as $n$ increases, while augmented Gibbs degrades as $n$ increases, as evidenced by increased autocorrelation.
  \item The merge cost $\omega$ again appears linear in $n$, further supporting our theory.
  \item Iteration time for both algorithms appears linear in $n$.
\end{itemize}

With partially collapsed mixing and iteration time remaining stable, we find a time per effective sample of order $n$. Conversely, the time per effective sample for the augmented Gibbs appears approximately quadratic, due to worsening mixing - in the language of \cite{liu1994fraction}, the augmented model exhibits an increase in the \emph{fraction of missing information}, in that $\pi(\theta | \psi, x)$ concentrates more quickly than $\pi(\theta | x)$. Hence, marginalizing out $\psi$ in the $\theta$-update results in empirically better scaling, rather than just a constant improvement. 

\subsubsection{Real data analysis: spatial distribution of a tropical forest tree}

Our second example concerns a central question in tropical-forest ecology: what governs where a species occurs, and whether community assembly reflects niche partitioning (species tracking distinct micro-environments) or neutral dynamics. The discriminating quantity is the species' spatial intensity, the expected stem density as a function of location: an intensity aligned with environmental structure signals habitat filtering, and estimating it underpins habitat mapping, climate-response forecasting and conservation planning. We consider the previously described GP-driven level-set Cox process with three levels. As data we use all $n=9{,}113$ living \emph{Alseis blackiana} stems from the 2015 census of the 50-ha Barro Colorado Island plot \citep{conditBCI2019}; rescaling both coordinates by a common factor preserves the true $2\!:\!1$ geometry on the window $[0,82.36] \times [0,41.18]$. The pattern shows broad, multi-level intensity variation, each level realised over several disconnected regions --- a genuine testbed for recovering a spatially disconnected, multi-level intensity surface. Finally, the size and complexity of this dataset motivate our methodological choice. Even in the deliberately simple synthetic study presented above, with two levels splitting the square into equal halves, a fully pseudo-marginal MCMC mixed increasingly poorly as the number of points grew. The present example is far more demanding, with $n=9{,}113$ points and three levels supported on spatially disconnected regions; a fully pseudo-marginal analysis is impractical here, whereas the approach developed in this paper remains feasible at this scale.

We run the collapsed Gibbs sampler previously described, updating the three intensity levels $\lambda_{1}, \lambda_{2}, \lambda_3$ in separate blocks and fixing at $-0.2$ and $1$ the two thresholds that partition the latent Gaussian field into levels. The uniform random-walk proposals are tuned to the optimal-scaling regime \citep{agrawal2023optimal}, giving acceptance rates around $0.3$; the Portkey truncation fires in about $8\%$ of iterations for each level. The chain is run for 10{,}000 iterations after a warm-up.

The C2C acceptance is parallelised across the $64$ tree leaves. Run on a workstation with a 52-thread processor, this yields a substantial speed-up: the average time to perform one update of each of the three levels $\lambda_{1},\lambda_{2},\lambda_3$ is around 5 times shorter than in the serial implementation. The gain would be larger still with more leaves. At each step of the C2C the 2-coin algorithm is parallelised only over the leaves that still require an update, and this number is typically (often considerably) smaller than the 52 available threads, so part of the processor sits idle.

\begin{figure}[!ht]
  \centering
  \includegraphics[width=1\linewidth]{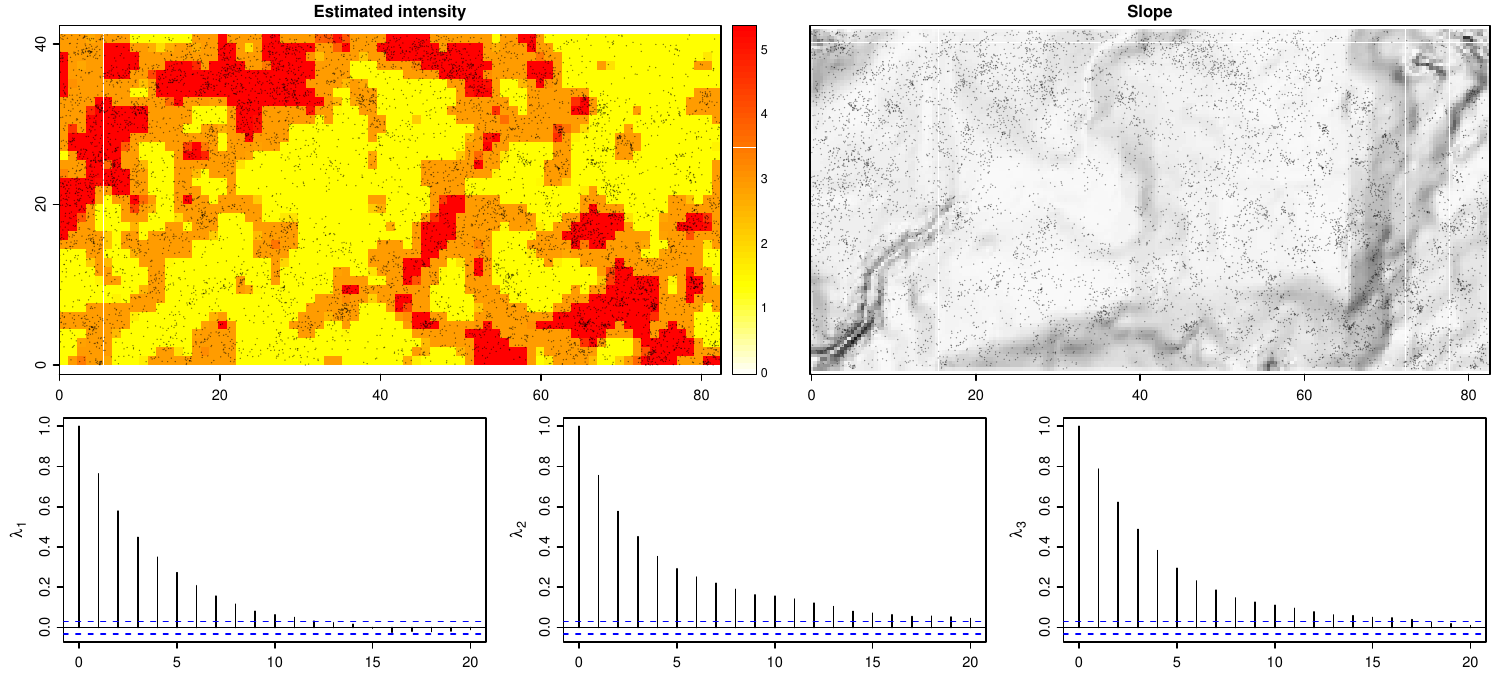}
  \caption{[Cox process application] Barro Colorado Island \emph{Alseis blackiana} example.
  \emph{Top:} (left) posterior-mean intensity surface estimated by the three-level model and (right) topographic slope (in degrees), both with the $n=9{,}113$ stem locations overlaid. \emph{Bottom:} autocorrelation functions
  of the three intensity levels $\lambda_{1}$, $\lambda_{2}$ and $\lambda_3$.}
  \label{fig:bci}
\end{figure}

Figure~\ref{fig:bci} shows the estimated intensity surface and the topographic slope with the stem locations overlaid, together with the autocorrelation plots of the three levels. The posterior means (standard deviations) are $\lambda_{1} = 1.40\,(0.032)$, $\lambda_{2} = 2.99\,(0.048)$ and $\lambda_3 = 5.50\,(0.103)$. Mixing is good, with effective sample sizes of 1342, 987 and 1190 for $\lambda_{1}$, $\lambda_{2}$ and $\lambda_3$, respectively. We compare the estimated intensity with topographic slope, the micro-environmental variable most directly linked to this species' documented habitat affinity (see Figure~\ref{fig:bci}).  High-intensity regions coincide broadly with steeper, sloping ground and are largely absent from the flat plateau interior, consistent with the reported preference of \emph{A.~blackiana} for slopes. The correspondence is, however, only partial and spatially uneven, indicating that slope alone does not govern the intensity and that additional, unmeasured factors also shape it. This is precisely the setting the nonparametric level-set model is built for: it recovers the intensity surface without assuming it to be a function of any measured covariate, so structure beyond slope is captured rather than forced away.

\section{Discussion}
\label{sec:discussion}

We have described a general Bernoulli factory MCMC approach to Bayesian inference for intractable likelihood models in large data scenarios. Through analysis and computer experiments, we have shown that these algorithms achieve polynomial scaling where previous Bernoulli MCMC implementations scaled exponentially in data size. 

In particular, we have shown sub-quadratic scaling in models where the likelihood is an intractable Girsanov-type exponentiated integral; to our knowledge this is the first such solution with documented sub-quadratic scaling. This theme suggests that whenever the full-conditional density of the parameters has the form $\pi(\theta, x) = \eta(\int_{S} \phi(z_{s}, y, \theta) \dd{s})$, for a non-negative, possibly non-linear link $\eta$, a generic C2C sampler can be constructed. A suitable partition $\braces{S_i}_{i=1}^{n}$ of the domain $S$ is defined to meet Assumption~\ref{ass:dec}. Based on the unbiased estimator $\op{E}[u \sim \op{Unif}{S_{i}}]{\op{Vol}(S_i) \phi(z_u, y, \theta)} = \int_{S_i} \phi$, Taylor series stochastic truncation techniques can be used to obtain an unbiased estimator of $\pi(\theta, x)$ for a broad class of functions $\eta$ \citep{chopin2025towards}. This recipe covers SDE-driven models, point processes, GP-modulated survival and hazard models, spatio-temporal fusion, and continuous-time Markov jump processes with time-varying intensities. A notable special case is $\eta(u) = u^{-1}$ which, applied to a (Taylor-series-based) unbiased estimator of $\pi(\theta, x)$, furnishes the unbiased estimator of $\pi^{-1}$ required by the flipped 2-coin variant of the C2C.

Beyond our emphasis on parameter inference, the C2C can also be effective for updating intractable latent variables $z$, such as infinite-dimensional processes, with its advantages depending on the context. If $z$ can be decomposed into conditionally independent components where an expensive but exact sampling method (e.g.\ rejection sampling) is available, the C2C may provide a more efficient solution. Conversely, when $z$ lacks such structure, C2C may offer a better trade-off between mixing efficiency and cost per iteration compared to non-marginal algorithms. However, the infinite-dimensionality of $z$ may hinder the use of the 2-coin algorithm in terms of posterior odds, requiring additional efforts to design an efficient alternative, such as using different partitions $\#_{k}$ for current and proposed values.

In spite of the improvements that we bring to the Bernoulli MCMC framework, it also bears pointing out its persistent limitations: our methodology does not extend to dense latent variable models of the form $\int \prod_{i=1}^{n} \pi(x_{i} | \theta, \psi) \pi(\psi | \theta) \dd{\psi}$, where marginal algorithms could provide great mixing benefits. In this class, a pseudo-marginal algorithm in quadratic time remains the leading, if imperfect solution.

\section*{Data availability}

Replication code for the computer experiments in Section~\ref{sec:exp} is hosted at \url{https://github.com/timsf/scale-bernoulli-mcmc}.

\section*{Acknowledgements}

The authors would like to thank Krzysztof Łatuszyński and Gareth Roberts for helpful discussions.
Timothée Stumpf-Fétizon has received funding from the European Research Council (ERC, PrSc-HDBayLe, Grant agreement No. 101076564). 
Flávio Gonçalves is funded by FAPEMIG - grant APQ-03830-26, and CNPq - grant 308536/2023-1.

\printbibliography
\appendix

\section{Technical Appendices for Simulation Experiments}
\label{app:sim}

\subsection{Application to Diffusion Models}
\label{app:diff}

For the update to $\pi(\theta | x)$, we use the factorization
\begin{equation}
  f_{i}(\theta, \vartheta) = \frac{\pi(x_{(t_{i-1}, t_{i}]} | \vartheta, x_{t_{i-1}}) \pi(\vartheta)^{1/n}}{\pi(x_{(t_{i-1}, t_{i}]} | \theta, x_{t_{i-1}}) \pi(\theta)^{1/n}} = \frac{d_{i}(\theta, \vartheta) q_{i}(\theta, \vartheta)}{d_{i}(\vartheta, \theta) q_{i}(\vartheta, \theta)},
\end{equation}
where $d_{i}$ is a tractable function, and
\begin{equation}
  q_{i}(\theta, \vartheta)
  = \op{exp}{-\int_{t_{i-1}}^{t_{i}} 0 \lor \braces{\varphi_{\vartheta} - \varphi_{\theta}}(x_{s}) \dd{s}} \in [0, 1].
\end{equation}
Coins with probability $q_{i}(\theta, \vartheta)$ can be simulated with the Poisson coin algorithm, which is a special case of the estimator in Example \ref{ex:poissonest}, and widely deployed in the exact diffusion inference literature.


\subsection{Application to Cox Processes}
\label{app:cox}

Noting that most $T_{i}$ lie almost entirely within a single $S_{l}$, we can optimize the C2C by using a \emph{flipped} 2-coin algorithm for some of the terms. The resulting factorization of the join density
\begin{gather}
  \frac{\pi(y \cap S_{l}, \vartheta_{l} | z \cap S_{l})}{\pi(y \cap S_{l}, \theta_{l} | z \cap S_{l})} = \prod_{i=1}^{n} \begin{dcases} \frac{\tilde{d}_{i}(\theta, \vartheta) \tilde{q}_{i}(\theta, \vartheta)}{\tilde{d}_{i}(\vartheta, \theta) \tilde{q}_{i}(\vartheta, \theta)} & \text{if $\op{centroid}{T_{i}} \in S_{l}$} \\ \frac{d_{i}(\theta, \vartheta) q_{i}(\theta, \vartheta)}{d_{i}(\vartheta, \theta) q_{i}(\vartheta, \theta)} & \text{else} \end{dcases}
\end{gather}
where $d_{i}, \tilde{d}_{i}$ are tractable functions, and $q_{i}, \tilde{q}_{i}$ are intractable probabilities given by
\begin{equation}
  \tilde{q}_{i}(\theta, \vartheta) = e^{-\int_{S} \op{max}{0, \theta_{l} - \vartheta_{l}} I_{T_{i} \setminus S_{l}}(s) \dd{s}}, \quad
  q_{i}(\theta, \vartheta) = e^{-\int_{S} \op{max}{0, \vartheta_{l} - \theta_{l}} I_{T_{i} \cap S_{l}}(s) \dd{s}}.
\end{equation}
Both probabilities are within $[e^{-|\vartheta_{l} - \theta_{l}|}, 1]$ by construction, so for a proposal $\vartheta_{l} | \theta_{l} \sim \op{Unif}{\theta_{l} \pm \delta / \sqrt{n}}$, the coins are lower bounded by $e^{-\delta / \sqrt{n}}$. Coins with probability $q_{i}(\theta, \vartheta)$ can be simulated with the Poisson estimator in Example \ref{ex:poissonest}.

We compare the benchmark algorithm---an augmented Gibbs sampling (AGS) with Metropolis steps---to a partially collapsed Gibbs sampler (CGS) that updates $z$ and $\psi$ according to their full conditionals via Metropolis-within-Gibbs, while $\theta$ is sampled from $\pi(\theta | z, y)$ using a Barker-within-Gibbs step. Hence, we integrate over $\psi$, rather than conditioning as in the benchmark, which typically results in better mixing. Note that the $\theta_{l}$'s are conditionally independent of each other, allowing for independent updates. The likelihood is factorized over a partition of $S$ into regular squares $\braces{T_{i}: i = 1, 2, \dots, n}$, where $n = \mu(S)$ is the natural data size index. 

The benchmark AGS was proposed in \citep{gonccalves2023cexact} to address the intractable likelihood arising from unknown level set areas. The augmented variable is a Poisson process $\psi$ on $S$ with rate $\nu(\ub{\lambda} - \lb{\lambda})$, where $\ub{\lambda} = \max \theta$, $\lb{\lambda} = \min \theta$ and $\nu > 1$. The estimator of $\exp\braces{-\int_{S} \lambda_{\theta}(z_{s}) \dd{s}}$ is given by
\begin{equation}
\label{eq:pmest}
  e^{-\mu(S)\lb{\lambda}} \prod_{l=1}^{L} \parens*{ \frac{\nu\ub{\lambda} - \theta_{l}}{\nu\ub{\lambda} - \lb{\lambda}}}^{|\psi \cap S_{l}|},
\end{equation}
where $\mu(S)$ is the area of $S$. The authors decouple the $\theta_{l}$ parameters from the distribution of $\psi$ such that the parameters $\theta$, $z$ and $\psi$ can be sampled in separate blocks via Metropolis-within-Gibbs. $\theta$ and $z$ follow random walk-type proposals while $\psi$ follows the independence proposal described above.

\section{Proofs of Main Text Propositions}
\label{sec:proofs}



\subsection{Proof of Proposition \ref{prop:c2c}}

We show, by induction on the height $\ell - k$ of the node $j_{1}\dots j_{k}$, that \text{C2C}$(\class{F}_{j_{1}\dots j_{k}})$ outputs $1$ with probability $r_{j_{1}\dots j_{k}}(\theta, \vartheta)$.

At a leaf ($k = \ell$), \text{C2C} returns \text{2C}$(\class{F}_{j_{1}\dots j_{\ell}})$, whose output probability is $r_{j_{1}\dots j_{\ell}}(\theta, \vartheta)$ by the validity of the 2-coin
algorithm.

For $k < \ell$, assume the claim for the two children. By Algorithm \ref{alg:c2c}, the call draws $C_{1}$ and $C_{2}$ from \text{C2C}$(\class{F}_{j_{1}\dots j_{k},1})$ and \text{C2C}$(\class{F}_{j_{1}\dots j_{k},2})$, which by the induction hypothesis are coins of probability $r_{j_{1}\dots j_{k},1}(\theta, \vartheta)$ and $r_{j_{1}\dots j_{k},2}(\theta, \vartheta)$. The call is therefore the 2-coin algorithm applied to these two coins, and by \eqref{eq:merge} its output probability is $r_{j_{1}\dots j_{k}}(\theta, \vartheta)$.

Taking $k = 0$ yields the marginal output probability $r_{0}(\theta, \vartheta)$.

\subsection{Proof of Proposition \ref{prop:c2ccost}}
\label{sec:c2ccost}

\begin{proof}
  We begin by noting that the cost at each node is an independent random variable by Assumption \ref{ass:dec}. For the base case at node $j_{1} \dots j_{\ell}$, the expected cost is $\rho_{j_{1} \dots j_{\ell}}$. Define $\rho_{j_{1} \dots j_{k}}$ for $k<l$, as the expected cost of tossing a $\rho_{j_{1} \dots j_{k}}$-coin and $\tau_{j_{1} \dots j_{k}}$ as the ENL of that task. Each of those merge loops requires outputs from the two children of node $j_{1} \dots j_{k}$, that is $j_{1} \dots j_{k}, 1$ and $j_{1} \dots j_{k}, 2$. Therefore, $\rho_{j_{1} \dots j_{k}} =(\rho_{j_{1} \dots j_{k}, 1} + \rho_{j_{1} \dots j_{k}, 2}) \tau_{j_{1} \dots j_{k}}$. Iterating up from the root, we find the C2C expected cost
  \begin{equation}
    \begin{aligned}
      \rho_{0}
      & = \tau_{0} (\rho_{1} + \rho_{2}) \\
      & = \tau_{0} ((\rho_{1,1} + \rho_{1,2}) \tau_{1} + (\rho_{2,1} + \rho_{2,2}) \tau_{2}) \\
      & = \tau_{0} \sum_{j_{1} \dots j_{\ell} \in \braces{1, 2}^{\ell}} \rho_{j_{1} \dots j_{\ell}} \prod_{k=1}^{\ell - 1} \tau_{j_{1} \dots j_{k}}. 
    \end{aligned}
  \end{equation}
\end{proof}

\subsection{Proof of Proposition \ref{prop:portkeydnc}}
\label{sec:portkeydnc}

\begin{proof}
  We proceed by recursion, and omit the proposal kernel for brevity of notation. The base case corresponds to the coins at the leaves $j_{1} \dots j_{\ell}$, which by \citep[Theorem 2]{vats2022efficient} has event probabilities
  \begin{equation}
    \frac{1}{b_{j_{1} \dots j_{\ell}}(\theta, \vartheta) + r_{j_{1} \dots j_{\ell}}(\theta, \vartheta) + r_{j_{1} \dots j_{\ell}}(\vartheta, \theta)} \times \begin{dcases} b_{j_{1} \dots j_{\ell}}(\theta, \vartheta)  & (\text{escape}) \\ r_{j_{1} \dots j_{\ell}}(\vartheta, \theta) & (\text{return 0}) \\ r_{j_{1} \dots j_{\ell}}(\theta, \vartheta) & (\text{return 1}) \end{dcases}.
  \end{equation}
  where $b_{j_{1} \dots j_{\ell}}(\theta, \vartheta) = b_{j_{1} \dots j_{\ell}}(\vartheta, \theta)$. We can recursively apply Lemma \ref{lem:portkeymerge} at node $j_{1} \dots j_{k}$ which merges nodes $j_{1} \dots j_{k}, 1$ and $j_{1} \dots j_{k}, 2$, finding that it has event probabilities
  \begin{equation}
    \frac{1}{b_{j_{1} \dots j_{k}}(\theta, \vartheta) + r_{j_{1} \dots j_{k}}(\theta, \vartheta) + r_{j_{1} \dots j_{k}}(\vartheta, \theta)} \times \begin{dcases} b_{j_{1} \dots j_{k}}(\theta, \vartheta)  & (\text{escape}) \\ r_{j_{1} \dots j_{k}}(\vartheta, \theta) & (\text{return 0}) \\ r_{j_{1} \dots j_{k}}(\theta, \vartheta) & (\text{return 1}) \end{dcases},
  \end{equation}
  where $b_{j_{1} \dots j_{k}}(\theta, \vartheta) = b_{j_{1} \dots j_{k}}(\vartheta, \theta)$, so nodes at level $\ell - 1$ have escape probability invariant to the direction of the MCMC move. Iterating all the way to the root, the portkey C2C has event probabilities
  \begin{equation}
    \frac{1}{b_{0}(\theta, \vartheta) + r_{0}(\theta, \vartheta) + r_{0}(\vartheta, \theta)} \times \begin{dcases} r_{0}(\vartheta, \theta) + b_{0}(\theta, \vartheta) & (\text{return 0}) \\ r_{0}(\theta, \vartheta) & (\text{return 1}) \end{dcases},
  \end{equation}
  where $b_{0}(\theta, \vartheta) = b_{0}(\vartheta, \theta)$. Since $r_{0}(\theta, \vartheta) = \pi(\vartheta, x) / (\pi(\vartheta, x) + \pi(\theta, x))$,
  \begin{equation}
    \frac{1}{\tilde{b}_{0}(\theta, \vartheta) + \pi(\vartheta, x) + \pi(\theta, x)} \times \begin{dcases} \pi(\theta, x) + \tilde{b}_{0}(\theta, \vartheta) & (\text{return 0}) \\ \pi(\vartheta, x) & (\text{return 1}) \end{dcases},
  \end{equation}
  where $\tilde{b}_{0}(\theta, \vartheta) = b_{0}(\theta, \vartheta) (\pi(\vartheta, x) + \pi(\theta, x))$ is invariant to parameter inversions. Therefore, the portkey C2C accepts the proposal with probability $\pi(\vartheta, x) / (\tilde{b}_{0}(\theta, \vartheta) + \pi(\vartheta, x) + \pi(\theta, x))$.
\end{proof}

\subsection{Proof of Proposition \ref{prop:balscale}}
\label{sec:balscale}

\begin{proof}
  In the first instance, since $h_{j_{1} \dots j_{k}}(\theta, \vartheta) = h_{j_{1} \dots j_{k}}(\theta, \theta) = 1$, $r_{j_{1} \dots j_{k}}(\theta, \vartheta) = 1/2$ and $\tau_{j_{1} \dots j_{k}} = 2$ at all nodes. The merge overhead simplifies to $\omega = \sum_{j_{1} \dots j_{\ell} \in \braces{1, 2}^{\ell}} 2^{\ell} = 4^{\ell}$ as $\braces{1, 2}^{\ell}$ has $2^{\ell}$ elements. In the second instance, we observe that if all $h_{j_{1} \dots j_{\ell}}$ are identical, then for all levels $k$, all $h_{j_{1} \dots j_{k}}$ are identical as well.  Then, at any node,
  \begin{equation}
     \begin{aligned}
       \tau_{j_{1} \dots j_{k}}
       & = \frac{r_{j_{1} \dots j_{k}}(\theta, \vartheta)}{r_{j_{1} \dots j_{k},1}(\theta, \vartheta) r_{j_{1} \dots j_{k},2}(\theta, \vartheta)} \\
       & = \frac{(1 + h_{j_{1} \dots j_{k},1}(\vartheta, \theta))(1 + h_{j_{1} \dots j_{k},2}(\vartheta, \theta))}{1 + h_{j_{1} \dots j_{k}}(\vartheta, \theta)} \\
       & = \frac{(1 + h_{j_{1} \dots j_{k},1}(\vartheta, \theta))^{2}}{1 + h_{j_{1} \dots j_{k},1}(\vartheta, \theta)^{2}} \\
       & = 1 + \frac{2 h_{j_{1} \dots j_{k},1}(\vartheta, \theta)}{1 + h_{j_{1} \dots j_{k},1}(\vartheta, \theta)^{2}} \\
       & \leq 2,
    \end{aligned}
  \end{equation}
  since $2ab \leq a^{2} + b^{2}$ for any $a, b$. It follows that $\omega \leq \sum_{j_{1} \dots j_{\ell} \in \braces{1, 2}^{\ell}} 2^{\ell} = 4^{\ell}$.
\end{proof}

\subsection{Proof of Proposition \ref{prop:morand}}
\label{sec:morand}

\begin{proof}
  To avoid notational clutter due to non-equal splits in the tree, we make the statements below for $(n, \ell)$ such that $n / 2^{\ell} \in \mathbf{N}$. We begin by noting that under uniformly random partitions, all $\prod_{k=1}^{\ell-1} \tau_{j_{1} \dots j_{k}}$ have identical distribution. Therefore,
  \begin{equation}
    \begin{aligned}
      2^{-\ell} \op{E}[\varsigma]{\omega}
      & = 2^{-\ell} \op{E}[\varsigma]{\sum_{j_{1} \dots j_{\ell} \in \braces{1, 2}^{\ell}} \tau_{0} \prod_{k=1}^{\ell-1} \tau_{j_{1} \dots j_{k}}} \\
      & = \op{E}[\varsigma]{\tau_{0} \prod_{k=1}^{\ell-1} \tau_{j_{1} \dots j_{k}}}, \qquad (j_{1} \dots j_{\ell} \in \braces{1, 2}^{\ell}),
    \end{aligned}
  \end{equation}
  and we can analyze the batch costs individually. Applying Proposition \ref{prop:c2ccost} and multiplying over all levels other than the leaves, we observe the cancellation
    \begin{equation}
    \begin{aligned}
      \tau_{0} \prod_{k=1}^{\ell - 1} \tau_{j_{1} \dots j_{k}}
      & = \frac{r_{0}(\vartheta, \theta)}{r_{1}(\vartheta, \theta) r_{2}(\vartheta, \theta)} \prod_{k=1}^{\ell - 1} \frac{r_{j_{1} \dots j_{k}}(\vartheta, \theta)}{r_{j_{1} \dots j_{k}, 1}(\vartheta, \theta) r_{j_{1} \dots j_{k}, 2}(\vartheta, \theta)} \\
      & = \frac{r_{0}(\vartheta, \theta)}{r_{j_{1} \dots j_{\ell - 1}, 1}(\vartheta, \theta) r_{j_{1} \dots j_{\ell - 1}, 2}(\vartheta, \theta)} \prod_{k=1}^{\ell - 1}\frac{r_{j_{1} \dots j_{k}}(\vartheta, \theta)}{r_{j_{1} \dots j_{k - 1}, 1}(\vartheta, \theta) r_{j_{1} \dots j_{k - 1}, 2}(\vartheta, \theta)} \\
      & = \frac{r_{0}(\vartheta, \theta)}{r_{j_{1} \dots j_{\ell}}(\vartheta, \theta) r_{j_{1} \dots j_{\ell - 1}, (3 - j_{\ell})}(\vartheta, \theta)} \prod_{k=1}^{\ell - 1}\frac{r_{j_{1} \dots j_{k}}(\vartheta, \theta)}{r_{j_{1} \dots j_{k}}(\vartheta, \theta) r_{j_{1} \dots j_{k - 1}, (3 - j_{k})}(\vartheta, \theta)} \\
      & = \frac{r_{0}(\vartheta, \theta)}{r_{j_{1} \dots j_{\ell}}(\vartheta, \theta) \prod_{k=1}^{\ell} r_{j_{1} \dots j_{k - 1}, (3 - j_{k})}(\vartheta, \theta)} \\
      & = r_{0}(\vartheta, \theta) \parens*{1 + h_{j_{1} \dots j_{\ell}}(\theta, \vartheta)} \prod_{k=1}^{\ell} \parens*{1 + h_{j_{1} \dots j_{k-1} (3 - j_{k})}(\theta, \vartheta)},
    \end{aligned}
  \end{equation}
  where we note that the products are driven by disjointed batches of factors. Therefore, if we expand the product, take expectations and apply exchangeability, we obtain
  \begin{equation}
    \begin{aligned}
    \op{E}[\varsigma]{\tau_{0} \prod_{k=1}^{\ell - 1} \tau_{j_{1} \dots j_{k}}}
    & = r_{0}(\vartheta, \theta) \sum_{k=1}^{2^{\ell + 1}} \op{E}[\varsigma]{\prod_{i \in I_{k}} f_{\varsigma(i)}(\theta, \vartheta)} \\
    & = r_{0}(\vartheta, \theta) \sum_{k=1}^{2^{\ell + 1}} \op{E}[\varsigma]{\prod_{i = 1}^{|I_{k}|} f_{\varsigma(i)}(\theta, \vartheta)},
    \end{aligned}
  \end{equation}
  where $I_{k}$ is the set of indices involved in the $k$th summand. Consequently, it is sufficient to find the factor counts. Those are given by $(0, n, 2^{-\ell}n, 2^{-\ell}n, 2 \times 2^{1-\ell}n, 2 \times 2^{1-\ell}n, \dots, (2^{\ell - 1}) 2^{\ell}n, (2^{\ell - 1}) 2^{\ell}n)$. Therefore,
 \begin{equation}
    \begin{aligned}
      \op{E}[\varsigma]{\tau_{0} \prod_{k=1}^{\ell-1} \tau_{j_{1} \dots j_{k}}}
      & = r_{0}(\vartheta, \theta) \parens*{1 + \prod_{i=1}^{n} f_{i}(\theta, \vartheta) + 2 \sum_{j=1}^{2^{\ell}-1} \op{E}[\varsigma]{\prod_{i=1}^{n j 2^{-\ell}} f_{\varsigma(i)}(\theta, \vartheta)}} \\
      & = 1 + 2 r_{0}(\vartheta, \theta) \sum_{j=1}^{2^{\ell}-1} \op{E}[\varsigma]{\prod_{i=1}^{n j 2^{-\ell}} f_{\varsigma(i)}(\theta, \vartheta)}.
    \end{aligned}
  \end{equation}
\end{proof}

\subsection{Proof of Proposition \ref{prop:leafenl}}
\label{sec:leafenl}

\begin{proof}
  We begin by deploying the same device as in Theorem \ref{thm:gibbsscale} and carry out the analysis under the asymptotic measure $\meas{H}_{n}$. The ENL $\tau_{j_{1} \dots j_{\ell}}$ is given by
  \begin{equation}
    \tau_{j_{1} \dots j_{\ell}} = \prod_{\braces{i: \#_{\ell}(i) = j_{1} \dots j_{\ell}}} q_{i}(\theta_{n}, \vartheta_{n})^{-1} = \prod_{\braces{i: \#_{\ell}(i) = j_{1} \dots j_{\ell}}} \sup_{\psi} \frac{\pi(\psi | x_{i}, \vartheta_{n})}{\pi(\psi | x_{i}, \theta_{n})}.
  \end{equation}
  Since data is IID by assumption by Assumption \ref{ass:dgp}, the expectation under $\meas{H}_{n}$ for fixed $(\theta_{n}, \vartheta_{n})$ factorizes to
  \begin{equation}
    \op{E}[\meas{H}_{n}]{\tau_{j_{1} \dots j_{\ell}} | \theta_{n}, \vartheta_{n}} = \op{E}[\meas{H}_{n}]{\sup_{\psi} \frac{\pi(\psi | x_{1}, \vartheta_{n})}{\pi(\psi | x_{1}, \theta_{n})} | \theta_{n}, \vartheta_{n}}^{2^{-\ell} n}.
  \end{equation}
  Assuming that $\class{E} = \braces{\braces{\theta_{n}, \vartheta_{n}} \subset \class{B}}$ applies, we expand the log integrand to first order around $\theta_{n}$, which yields
  \begin{equation}
    \log \frac{\pi(\psi | x_{1}, \vartheta_{n})}{\pi(\psi | x_{1}, \theta_{n})} = |\vartheta_{n} - \theta_{n}| \partial_{\dot{\theta}} \log \pi(\psi | x_{1}, \dot{\theta})
    \leq \frac{\delta}{\sqrt{n}} \sup_{\theta \in \class{B}} |\partial_{\theta} \log \pi(\psi | x_{1}, \theta)|
  \end{equation}
  for some $\dot{\delta} \in \class{B}$. Making $n$ as large as necessary to apply the MGF bound, by elementary properties of subexponential RVs there is a $\zeta(\class{B}) > 0$ such that
  \begin{equation}
    \op{E}[\meas{H}_{n}]{\exp\braces*{\frac{\delta}{\sqrt{n}} \sup_{\theta \in \class{B}, \psi} |\partial_{\theta} \log \pi(\psi | x_{1}, \theta)|} | \theta_{n}, \vartheta_{n}} \leq \exp\braces*{\frac{\delta}{\sqrt{n}} \zeta(\class{B})}
  \end{equation}
  for large enough $n$ and on $\class{E}$. Thus,
  \begin{equation}
    \op{E}[\meas{H}_{n}]{\tau_{j_{1} \dots j_{\ell}} | \theta_{n}, \vartheta_{n}} \leq \exp\braces*{\delta \zeta(\class{B})}^{2^{-\ell} n}
  \end{equation}
  for large enough $n$ and on $\class{E}$. The bound being flat over $\class{E}$, $\op{E}[\meas{H}_{n}]{\tau_{j_{1} \dots j_{\ell}} | \class{E}} \leq \exp\braces*{\delta \zeta(\class{B})}^{2^{-\ell} n}$ as well by the tower property. The rest of the proof proceeds as in Theorem \ref{thm:gibbsscale}.
\end{proof}

\subsection{Proof of Theorem \ref{thm:gibbsscale}}
\label{sec:gibbsscale}

\begin{proof}
  To begin with, we observe that by Assumption \ref{ass:post}, the law of $(\theta_{n} | y_{1:n})$ converges to $N[\theta_{0}, I_{0}^{-1}]$ in total variation. Therefore, we can derive the result under the more simple measure $\meas{H}_{n}$ with density
  \begin{equation}
    \pi(y_{1:n} | \theta_{0}) \pi(z_{1:n} | \theta_{n}) N[\theta_{n}; \theta_{0}, I_{0}^{-1}] \op{Unif}{\vartheta_{n}; \theta_{n} \pm \delta / \sqrt{n}}
  \end{equation}
  and push it back to $\meas{G}_{n}$ in the limit. Moreover, due to posterior concentration, $\meas{H}_{n}(\braces{\theta_{n}, \vartheta_{n}} \subset \class{B})$ can be made arbitrarily close to 1 for large enough $n$. Therefore, we will work on the event $\class{E} = \braces{(\theta_{n}, \vartheta_{n}) \subset \class{B}}$. Applying Proposition \ref{prop:morand}, restricting the model to conditionally IID, and taking conditional expectations,
  \begin{equation}
    \begin{aligned}
      \op{E}[\meas{H}_{n}]{\omega_{n} | \theta_{n}, \vartheta_{n}}
      & = 2^{\ell} \parens*{1 + 2 r_{0}(\vartheta, \theta) \sum_{j=1}^{2^{\ell}-1} \op{E}[\meas{H}_{n}]{\frac{\pi(y_{1}, z_{1} | \vartheta_{n})}{\pi(y_{1}, z_{1} | \theta_{n})} | \theta_{n}, \vartheta_{n}}^{nj 2^{-\ell}}} \\
      & \leq 2^{\ell} \parens*{1 + 2 (2^{\ell}-1) \op{E}[\meas{H}_{n}]{\frac{\pi(y_{1}, z_{1} | \vartheta_{n})}{\pi(y_{1}, z_{1} | \theta_{n})} | \theta_{n}, \vartheta_{n}}^{n}}
    \end{aligned}
  \end{equation}
  Helpfully, the expected density ratio simplifies to
  \begin{equation}
    \begin{aligned}
      \op{E}[\meas{H}_{n}]{\frac{\pi(y_{1}, z_{1} | \vartheta_{n})}{\pi(y_{1}, z_{1} | \theta_{n})} | \theta_{n}, \vartheta_{n}} 
      & = \iint \frac{\pi(y_{1}, z_{1} | \vartheta_{n})}{\pi(y_{1}, z_{1} | \theta_{n})} \pi(z_{1} | y_{1}, \theta_{n}) \pi(y_{1} | \theta_{0}) \dd{z_{1}} \dd{y_{1}} \\
      & = \iint \pi(y_{1}, z_{1} | \vartheta_{n}) \dd{z_{1}} \frac{\pi(y_{1} | \theta_{0})}{\pi(y_{1} | \theta_{n})} \dd{y_{1}} \\
      & = \int \pi(y_{1} | \vartheta_{n}) \frac{\pi(y_{1} | \theta_{0})}{\pi(y_{1} | \theta_{n})} \dd{y_{1}} \\
      & = \op{E}[\meas{P}_{0}]{\frac{\pi(y_{1} | \vartheta_{n})}{\pi(y_{1} | \theta_{n})}},
    \end{aligned}
  \end{equation}
  so we can disregard the latent model. In particular, by Lemma \ref{lem:stat}, there is an $\gamma(\class{B}) < \infty$ such that
  \begin{equation}
    \begin{aligned}
        \op{E}[\meas{H}_{n}]{\frac{\pi(y_{1} | \vartheta_{n})}{\pi(y_{1} | \theta_{n})} | \class{E}} \leq 2 \gamma(\class{B})^{1/n}
    \end{aligned}
  \end{equation}
  for large enough $n$. It follows that
  \begin{equation}
    \op{E}[\meas{H}_{n}]{\omega_{n} | \class{E}}
    \leq 2^{\ell} \parens*{1 + 4 (2^{\ell}-1) \gamma(\class{B})}
  \end{equation}
  for large enough $n$, and $\limsup_{n, \ell \to \infty} 4^{-\ell} \op{E}[\meas{H}_{n}]{\omega_{n} | \class{E}} \leq 4 \gamma(\class{B})$. By Markov's inequality,
  \begin{equation}
      \meas{H}_{n}[4^{-\ell} \omega_{n} \geq a] = \frac{\meas{H}_{n}[4^{-\ell} \omega_{n} \geq a | \class{E}]}{\meas{H}_{n}[\class{E}]} \leq \frac{4^{-\ell} \op{E}[\meas{H}_{n}]{\omega_{n} | \class{E}}}{a \meas{H}_{n}[\class{E}]}, \qquad (a > 0)
  \end{equation}
  Observing that $\liminf_{n \to \infty} \meas{H}_{n}[\class{E}] = 1$ due to posterior concentration,
  \begin{equation}
      \limsup_{n, \ell \to \infty} \meas{H}_{n}[4^{-\ell} \omega_{n} \geq a] \leq \frac{4 \gamma(\class{B})}{a}.
  \end{equation}
  We now shift back to the original stationary measure $\class{G}_{n}$ by way of the total variation bound
  \begin{equation}
      \limsup_{n, \ell \to \infty} \meas{G}_{n}[4^{-\ell} \omega_{n} \geq a] \leq \lim_{n \to \infty} \abs{\meas{G}_{n} - \meas{H}_{n}}_{\mathrm{TV}} + \limsup_{n, \ell \to \infty} \meas{H}_{n}[4^{-\ell} \omega_{n} \geq a],
  \end{equation}
  and since $\meas{G}_{n} \to \meas{H}_{n}$ in total variation by Assumption \ref{ass:post},
  \begin{equation}
    \limsup_{n, \ell \to \infty} \meas{G}_{n}[4^{-\ell} \omega_{n} \geq a] \leq \frac{4 \gamma(\class{B})}{a}.
  \end{equation}
  We finally verify the elementary definition of stochastic boundedness,
  \begin{equation}
    \lim_{a \to \infty} \limsup_{n, \ell \to \infty} \meas{G}_{n}[4^{-\ell} \omega_{n} \geq a] = 0,
  \end{equation}
  and thus $4^{-\ell} \omega_{n} = O_{\meas{G}_{n}}(1)$.
\end{proof}

\section{Further Lemmas}
\label{sec:lemmata}

\begin{lemma}[Stationary likelihood ratio]
  \label{lem:stat}
   Suppose that Assumptions \ref{ass:dgp} and \ref{ass:smooth} apply, and $\theta_{n} = \theta_{0} + v / \sqrt{n}$ and $\vartheta_{n} = \theta_{n} + u / \sqrt{n}$ are both contained in $\class{B}$. Define $\class{E} = \braces{\braces{\theta_{n}, \vartheta_{n}} \subset \class{B}}$. Then, for $u \in [\pm \delta]$, $v \sim \op{N}{0, I_{0}^{-1}}$, and $n$ large enough, there is a constant $\gamma(\class{B}) > 0$ such that
  \begin{equation}
    \op{E}{\frac{\pi(y_{1} | \vartheta_{n})}{\pi(y_{1} | \theta_{n})} | \class{E}} \leq 2 \gamma(\class{B})^{1/n}.
  \end{equation}
  \begin{proof}
    For fixed $\braces{\theta_{n}, \vartheta_{n}} \subset \class{B}$ and $n$ large enough, Lemma \ref{lem:likrat} yields
    \begin{equation}
      \op{E}[\meas{P}_{0}]{\frac{\pi(y_{1} | \vartheta_{n})}{\pi(y_{1} | \theta_{n})}} \leq \exp\braces*{|v| u \iota(\class{B}) + u^{2} (2\xi(\class{B})^{2} +\zeta(\class{B}) / 2)}^{1/n},
    \end{equation}
    so we need to average over $u$ and $v$ on $\class{E}$. We naively bound $|u| \leq \delta$, and observe that $|v| \leq \sqrt{n} \sup_{a, b \in \class{B}} |a - b|$, i.e. $v$ has truncated Gaussian distribution. It follows that $\op{E}[v]{\exp\braces*{|v| \delta \iota(\class{B}) / n} | \class{E}}$ is an expectation with respect to a truncated lognormal random variable. Since the integrand increases in $|v|$, we can bound by dropping the truncation, apply $e^{a|v|} \leq e^{av} + e^{-av}$, and obtain
    \begin{equation}
      \begin{aligned}
        \op{E}[v]{\exp\braces*{|v| \delta \iota(\class{B}) / n} | \class{E}} 
        & \leq \op{E}[v]{\exp\braces*{|v| \delta \iota(\class{B}) / n}} \\
        & \leq \op{E}[v]{\exp\braces*{v \delta \iota(\class{B}) / n}} + \op{E}[v]{\exp\braces*{-v \delta \iota(\class{B}) / n}} \\
        & \leq 2 \exp\braces*{\delta^{2} I_{0}^{2} \iota(\class{B})^{2} / (2n)}.
      \end{aligned}
    \end{equation}
    In conjunction,
    \begin{equation}
      \op{E}{\frac{\pi(y_{1} | \vartheta_{n})}{\pi(y_{1} | \theta_{n})} | \class{E}} \leq 2 \exp\braces*{\delta^{2} (I_{0}^{2} \iota(\class{B})^{2} / 2  + 2\xi(\class{B})^{2} +\zeta(\class{B}) / 2)}^{1/n}
    \end{equation}
    for large enough $n$.
  \end{proof}
\end{lemma}

\begin{lemma}[Expected likelihood ratio]
  \label{lem:likrat}
  Suppose that Assumptions \ref{ass:dgp} and \ref{ass:smooth} apply, and $\theta_{n} = \theta_{0} + v / \sqrt{n}$ and $\vartheta_{n} = \theta_{n} + u / \sqrt{n}$ are both contained in $\class{B}$. Then, for large enough $n$, there are constants $\iota(\class{B}), \xi(\class{B}), \zeta(\class{B}) > 0$ such that
  \begin{equation}
    \op{E}[\meas{P}_{0}]{\frac{\pi(y_{1} | \vartheta_{n})}{\pi(y_{1} | \theta_{n})}} \leq \exp\braces*{|uv| \iota(\class{B}) + u^{2} (2\xi(\class{B})^{2} +\zeta(\class{B}) / 2)}^{1/n}.
  \end{equation}
  \begin{proof}
    In what follows, all expectations are with respect to $\meas{P}_{0}$. We expand $l(\vartheta_{n}) = \log \pi(y_{1} | \vartheta_{n})$ to second order around $\theta_{n}$, obtaining
    \begin{equation}
      \begin{aligned}
        \log \frac{\pi(y_{1} | \vartheta_{n})}{\pi(y_{1} | \theta_{n})}
        & = (\vartheta_{n} - \theta_{n}) l'(\theta_{n}) + R(y_{1}) \\
        & \leq \frac{u}{\sqrt{n}} l'(\theta_{n}) + \frac{u^{2}}{2n} \sup_{\theta \in \class{B}} |l''(\theta)|,
      \end{aligned}
    \end{equation}
    where $R$ is a remainder term. Going back to the original scale, substitute the expansion and apply Cauchy-Schwarz to obtain
    \begin{equation}
      \begin{aligned}
        \op{E}{\frac{\pi(y_{1} | \vartheta_{n})}{\pi(y_{1} | \theta_{n})}}
        & \leq \op{E}{\exp\braces*{\frac{u}{\sqrt{n}} l'(\theta_{n}) + \frac{u^{2}}{2n} \sup_{\theta \in \class{B}} |l''(\theta)|}} \\
        & \leq \exp\braces*{\frac{u}{\sqrt{n}} \op{E}{l'(\theta_{n})}} \op{E}{\exp\braces*{\frac{2u}{\sqrt{n}} (l'(\theta_{n}) - \op{E}{l'(\theta_{n}))}}}^{1/2} \\
        & \qquad \times \op{E}{\exp\braces*{\frac{u^{2}}{n} \sup_{\theta \in \class{B}} |l''(\theta)|}}^{1/2}.
      \end{aligned}
    \end{equation}
    For the first factor, we again expand the linear term around $\theta_{0}$ and take expectations, yielding
    \begin{equation}
      \begin{aligned}
        \op{E}{l'(\theta_{n})} 
        & = \op{E}{l'(\theta_{0}) + \int_{\theta_{n} \land \theta_{0}}^{\theta_{n} \lor \theta_{0}} l''(\theta) \dd{\theta}} \\
        & \leq |\theta_{n} - \theta_{0}| \sup_{\theta \in \class{B}} \op{E}{|l''(\theta)|} = \frac{|v|}{\sqrt{n}} \sup_{\theta \in \class{B}} \op{E}{|l''(\theta)|}.
      \end{aligned}
    \end{equation}
    where we've applied $\op{E}{l'(\theta_{0})} = 0$ by Assumption \ref{ass:smooth}, and notice that the expectation is finite by the same assumption. For the second factor, we observe that $l'(\theta_{n})$ is sub-exponential on $\class{B}$, and therefore, for large enough $n$, there is a $\xi(\class{B}) > 0$ such that 
    \begin{equation}
      \op{E}{\exp\braces*{\frac{2u}{\sqrt{n}} (l'(\theta_{n}) - \op{E}{l'(\theta_{n}))}}} \leq \exp\braces*{\frac{4u^{2}}{n} \xi(\class{B})^{2}}. \qquad (2|u|/\sqrt{n} \leq 1 / \xi(\class{B}))
    \end{equation}
    Note that $n$ can be made large enough for the bound to apply. Finally, for the last factor, $\sup_{\theta \in \class{B}} |l''(\theta)|$ is again sub-exponential, and there is a $\zeta(\class{B}) > 0$ such that we may apply the Bernstein-type bound
    \begin{equation}
      \op{E}{\exp\braces*{\frac{u^{2}}{n} \sup_{\theta \in \class{B}} |l''(\theta)|}} \leq \exp\braces*{\frac{u^{2}}{n} \zeta(\class{B})}. \qquad (u^{2}/n \leq 1 / \zeta(\class{B}))
    \end{equation}
    Assembling those bounds,
    \begin{equation}
      \op{E}{\frac{\pi(y_{1} | \vartheta_{n})}{\pi(y_{1} | \theta_{n})}}
      \leq \exp\braces*{|uv| \sup_{\theta \in \class{B}} \op{E}{|l''(\theta)|} + u^{2} (2\xi(\class{B})^{2} +\zeta(\class{B}) / 2)}^{1/n}
    \end{equation}
    for large enough $n$.
  \end{proof}
\end{lemma}

\begin{lemma}[Nested portkey algorithm]
  \label{lem:portkeymerge}
  Consider the superordinate portkey 2-coin algorithm whose 2-coins are themselves portkey 2-coin algorithms that escape with probability $1 - \varpi$ at every iteration, or in case its input coins escape, where the input coins have event probabilities
  \begin{equation}
    \frac{1}{b_{j}(\theta, \vartheta) + r_{j}(\vartheta, \theta) + r_{j}(\theta, \vartheta)} \times \begin{dcases} b_{j}(\theta, \vartheta)  & (\text{escape}) \\ r_{j}(\vartheta, \theta) & (\text{return 0}) \\ r_{j}(\theta, \vartheta) & (\text{return 1}) \end{dcases}, \qquad (j = 1, 2)
  \end{equation}
  and $b_{j}(\theta, \vartheta) = b_{j}(\vartheta, \theta)$. Then, the superordinate Portkey merge algorithm has the event probabilities
  \begin{equation}
    \frac{1}{b_{0}(\theta, \vartheta) + r_{0}(\vartheta, \theta) + r_{0}(\theta, \vartheta)} \times \begin{dcases} b_{0}(\theta, \vartheta) & (\text{escape}) \\ r_{0}(\vartheta, \theta) & (\text{return 0}) \\ r_{0}(\theta, \vartheta) & (\text{return 1}) \end{dcases},
  \end{equation}
  for some $b_{0}(\cdot, \cdot)$ such that $b_{0}(\theta, \vartheta) = b_{0}(\vartheta, \theta)$.
\begin{proof}
  Define $\eta(\theta, \vartheta) = \prod_{j = 1, 2} (b_{j}(\theta, \vartheta) + r_{j}(\vartheta, \theta) + r_{j}(\theta, \vartheta))$, which is invariant to the flipping of its arguments. Noting that at each iteration, the non-termination probability is
  \begin{equation}
    \varpi \parens*{1 - \frac{r_{0}(\vartheta, \theta) + r_{0}(\theta, \vartheta) + b_{1} + b_{2} - b_{1} b_{2}}{\eta}},
  \end{equation}
  the overall escape probability is
  \begin{equation}
    \begin{aligned}
      & \kern-1em ((1 - \varpi) + \varpi \frac{b_{1} + b_{2} - b_{1} b_{2}}{\eta} \sum_{k=0}^{\infty} \braces*{\varpi \parens*{1 - \frac{r_{0}(\vartheta, \theta) + r_{0}(\theta, \vartheta) + b_{1} + b_{2} - b_{1} b_{2}}{\eta}}}^{k} \\
      & = \frac{(1 - \varpi) + \varpi \frac{b_{1} + b_{2} - b_{1} b_{2}}{\eta}}{1 - \braces*{\varpi \parens*{1 - \frac{r_{0}(\vartheta, \theta) + r_{0}(\theta, \vartheta) + b_{1} + b_{2} - b_{1} b_{2}}{\eta}}}} \\
      & = \frac{(\varpi^{-1} - 1) \eta + b_{1} + b_{2} - b_{1} b_{2}}{(\varpi^{-1} - 1) \eta + b_{1} + b_{2} - b_{1} b_{2} + r_{0}(\vartheta, \theta) + r_{0}(\theta, \vartheta)},
      \end{aligned}
    \end{equation}
    where we identify
    \begin{equation}
      b_{0}(\theta, \vartheta) = (\varpi^{-1} - 1) \eta(\theta, \vartheta) + b_{1}(\theta, \vartheta) + b_{2}(\theta, \vartheta) - b_{1}(\theta, \vartheta) b_{2}(\theta, \vartheta),
    \end{equation}
    and observe its invariance to argument flips. By analogous arguments, we find the probability of returning 0 or 1.
  \end{proof}
\end{lemma}

\begin{proposition}[Exponential growth of the expected cost]
\label{prop:expE}
Let $\meas{P}_{n}$ be the measure with density 
\begin{equation}
  \pi(y_{1:n} | \theta_{0}) \pi(z_{1:n} | y_{1:n}, \theta_{n}),
\end{equation}
where $\theta_0$ is the true parameter value and $\theta_n$ is any value in the state space of $\theta$.

Define $r_i(\theta) \coloneqq -\log(p_i(\theta))$, $S_n = \sum_{i=1}^nr_i(\theta)$ and write $C(n) = \parens{\prod_{i=1}^np_i(\theta)}^{-1}$ as the lower bound on the ENL. Assume that $\exists \, \varepsilon > 0: \; \op{E}[\meas{P}_{n}]{r_i(\theta)} \ge \varepsilon, \; \forall \; i.$ Then, by Jensen's inequality
\begin{equation}
  \op{E}{C(n)} \geq e^{\varepsilon n}.
\end{equation}
\end{proposition}

\begin{proposition}[Exponential decay of tail probability]
\label{prop:expP}
Consider the same definitions and assumptions of Proposition \ref{prop:expE}. Define the natural filtration $\mathcal F_{i} \coloneqq \sigma(p_1(\theta), \dots, p_i(\theta))$ and the martingale $M_n \coloneqq S_n-\sum_{i=1}^{n} E_{\meas{P}_{i}}[r_i(\theta)\mid\mathcal F_{i-1}]$, with differences $\Delta_i \coloneqq M_i-M_{i-1}$ and predictable quadratic variation $V_n \coloneqq \sum_{i=1}^{n} \op{E}{\Delta_i^{2}\mid\mathcal{F}_{i-1}}$. Assume that there exist constants $K>0$ and $\lambda_{0} > 0$ such that
\begin{equation}
  \op{E}[\meas{P}_{i}]{e^{\lambda\Delta_i} | \mathcal F_{i-1}}
  \le \exp\braces*{\tfrac12 K^{2}\lambda^{2}} 
  \quad \meas{P}_{i-1}-\text{a.s.} \qquad (|\lambda| \le \lambda_{0}, i \geq 1).
\end{equation}
Then, $M_n$ obeys the following general Bernstein-type bound for martingale difference sequence \citep[Theorem~2.19]{wainwright2019high}:
\begin{equation}
  \op{\meas{P}_{\mathnormal{n}}}{S_n\le \op{E}[\meas{P}_{n}]{S_n} -\tfrac12\varepsilon n}
  \;\le\; 2e^{-\,c_{\varepsilon,K}\;n},
  \qquad c_{\varepsilon,K} \coloneqq \min\braces*{\frac{\varepsilon^2}{8K^2},\frac{\lambda_0\varepsilon}{4}}
\end{equation}
and the assumption $\op{E}[\meas{P}_{n}]{S_n} \ge \varepsilon n$ yields the exponential tail
\begin{equation}
  \op{\meas{P}_{\mathnormal{n}}}{\prod_{i=1}^np_i(\theta) \ge e^{-\varepsilon n/2}}
  \;\le\; 2e^{-c_{\varepsilon,K}\;n}.
\end{equation}
Therefore, $\op{E}[\mathbb{P}_{n}]{\prod_{i=1}^{n} p_{i}(\theta)} \;\le\; 2e^{-c_{\varepsilon,K}n} + e^{-\varepsilon n/2} \;\le\; 3 \rho^{n}$ for $\rho = \max\braces{e^{-c_{\varepsilon,K}},\,e^{-\varepsilon/2}} < 1$, and hence, $\op{E}[\mathbb{P}_{n}]{\prod_{i=1}^{n} p_{i}(\theta)} \;=\; O(\rho^{\,n})$.
\end{proposition}

Propositions \ref{prop:expE} and \ref{prop:expP} can be simplified if the $p_i$'s are independent and/or identically distributed \citep[Proposition~2.9]{wainwright2019high}.

\end{document}